\documentstyle[12pt,epic,eepic]{article}
\def\baselinestretch{1.1}
\renewcommand{\theequation}{\thesection.\arabic{equation}}
\newcommand{\newsection}{    
\setcounter{equation}{0}\section}
\newcommand{\be}{\nopagebreak\begin{equation}}
\newcommand{\ee}{\end{equation}}
\newcommand{\ba}{\begin{array}}
\newcommand{\ea}{\end{array}}
\newcommand{\bp}{\begin{picture}}
\newcommand{\ep}{\end{picture}}
\newcommand{\eol}{\\[1pc]}
\newcommand{\rf}[1]{Ref.~\cite{#1}}
\newcommand{\eq}[1]{Eq.~(\ref{#1})}
\newcommand{\bi}[6]{\bibitem{#1}{#2, }{\sl #3 }{\bf #4}{ (#5)}{ #6}}
\newcommand{\wt}[1]{\widetilde{#1}}
\newcommand{\wh}[1]{\widehat{#1}}
\newcommand{\ie}{{\em i.e., }}
\newcommand{\eg}{{\em e.g., }}
\renewcommand{\d}{\partial}
\renewcommand{\emptyset}{\mbox{\O}}
\newcommand{\eps}{\varepsilon}
\newcommand{\al}{{\cal A}}
\newcommand{\tr}{{\rm tr}}
\newcommand{\qtr}{{\rm qtr}}
\newcommand{\r}{\check{R}}
\newcommand{\scr}{\scriptstyle}
\newcommand{\scs}{\scriptscriptstyle}
\newcommand{\dsty}{\displaystyle}
\newcommand{\tsty}{\texstyle}
\newcommand{\R}{{\scriptscriptstyle R}}
\newcommand{\T}{{\scriptscriptstyle T}}
\newcommand{\q}{q=e^{i\frac{2\pi}{k+2}}}
\newcommand{\mv}[1]{\langle #1 \rangle}
\renewcommand{\thefootnote}{\fnsymbol{footnote}}
\newcommand{\M}{{\cal M}}
\newcommand{\I}{{\cal I}}
\newcommand{\Uq}{{\cal U}_q}
\newcommand{\Fq}{{\cal F}_q}
\newcommand{\so}{\sigma^0}
\newcommand{\si}{\sigma^1}
\newcommand{\sz}{\sigma^2}
\newcommand{\sm}{\sigma^3}

\begin{document}
\begin{titlepage}
\begin{flushright}
57/96/EP\\
April 1996
\end{flushright}
\vspace*{8pc}
\begin{center}
{\Huge \bf Quantum Deformation Of Lattice Gauge Theory}
\end{center}
\vspace{2pc}
\begin{center}
 {\Large D.V. Boulatov}\\
\vspace{1pc}
{\em International School for Advanced Studies (SISSA/ISAS)\\
via Beirut 2-4, I-34014 Trieste, Italy}
\vspace{2pc}
\end{center}
\begin{center}
{\large\bf Abstract}
\end{center}
A quantum deformation of 3-dimensional lattice gauge theory is defined by 
applying the Reshetikhin-Turaev functor to a Heegaard diagram associated to 
a given cell complex. In the root-of-unity case, the construction is
carried out with a modular Hopf algebra. In the topological
(weak-coupling) limit, the gauge theory partition function gives a 3-fold
invariant, coinciding in the simplicial case with the Turaev-Viro one. 
We discuss bounded manifolds as well as links in manifolds. By a 
dimensional reduction, we obtain a $q$-deformed gauge theory on Riemann
surfaces and find a connection with the algebraic Alekseev-Grosse-Schomerus
approach. 

\vfill
\end{titlepage}     

\stepcounter{footnote}

\bigskip     

\newsection{Introduction}

The lattice regularization of non-abelian gauge theory (LGT) proposed
by K.Wilson in 1975 \cite{Wilson} plays a fundamental role of a
non-perturbative definition of QCD. The main principle has been to
give up the Poincar\'{e} invariance of the theory and preserve the local
gauge symmetry as more fundamental. In the weak coupling regime, the
broken translational and rotational invariance restores dynamically.
If the gauge coupling is strictly equal to 0, the gauge strength
tensor vanishes identically and the theory becomes topological: one
can take any finite lattice from a given equivalence class without
changing the content of the model.

The lattice formulation extends in a natural way the set of acceptable
gauge groups for all compact groups while the continuous one is based on
the notion of the Lie algebra thus excluding finite groups, for
example. After the theory of quantum groups had appeared as a distinct
mathematical subject \cite{Qgroups,Woron}, the natural question arose
whether the notion of gauge symmetry could be extended to incorporate
quantum groups as well. This problem has not been only of academic
interest. As was noticed in \rf{B1}, the Ponzano-Regge model \cite{PR}
(which coincides with the classical ($q\to1$) limit of the Turaev-Viro
construction \cite{TV}) can be represented as LGT defined on lattices dual
to simplicial complexes. Then it was natural to assume that the Turaev-Viro
invariant could be related to a kind of LGT built on quantum group symmetry
(for the sake of brevity, we shall call it qQCD$_3$). This program was
carried out in \rf{B2} explicitely (see also \cite{B3}).  Technical
difficulties originating from a complicated structure of the representation
ring of $SL_q(2)$ at a root of unity, $\q$, was avoided in \cite{B2} by
establishing a direct connection with the ribbon graph invariants of
Reshetikhin and Turaev \cite{ReshTur}. The gauge invariance is implicit in
this formulation (however, it does not mean that the model does not enjoy
it). On the other hand, relative simplicity makes this explicit
representation very convenient.

The classical ($q\to1$) limit of the Turaev-Viro construction was
discovered long ago by Ponzano and Regge in the framework of Regge calculus
\cite{PR}. They argued that it can be regarded as a discretization of 3d
gravity with the Einstein-Hilbert action. On the other hand, the
Turaev-Viro invariant is related to Witten's Chern-Simons invariant
\cite{Wit}. Witten has shown that, with the ISO(2,1) gauge group, the
latter is connected with 3d quantum gravity. In the Euclidean regime for a
negative cosmological constant, the gauge group becomes isomorphic to
$SO(4)=SU(2)\times SU(2)$. It means that qQCD$_3$ possessing $SL_q(2)$
gauge symmetry is interesting from the physical point of view (an
exposition of the subject can be found in \rf{B3}).

The structure of the answers for the partition function suggests that, in 2
dimensions, the corresponding q-deformed LGT (qQCD$_2$) is related to the
topological $(G/G)_k$ coset model (as was argued in \rf{B4} these two
models are in some sense dual to each other).

An alternative purely algebraic approach to qQCD$_2$ was put forward in
\rf{A,AGS,BR}. The starting point for them was a Poisson bracket on 2d
lattice connections proposed by Fock and Rosly \cite{FR}. 

In 2 dimensions there is a cyclic order of links incident to a vertex.
Demanding that variables performing gauge rotations at each vertex form a
quasi-triangular Hopf algebra, Alekseev, Grosse and Schomerus have deduced
an algebra of gauge fields. In contrast to this situation, in 3 dimensions
there is a natural cyclic order of faces sharing the same link of a
lattice, which suggests that one should start with gauge fields forming a
Hopf algebra while gauge transformations are interpreted as changes of
bases these fields act on. As in the quantum case there is no group
manifold behind the construction, gauge fields and gauge transformations
have clearly the different statuses. One could say that this is one more
occurrence of the principle ``Quantization removes degenerations''.
Descending from 3 to 2 dimensions, one finds a model which seems, at first
sight, to be different from the AGS one. However, as we shall show they are
locally equivalent.

The outline of the paper is the following.\\
Chapter~2 is devoted to the construction of $q$-deformed
LGT in three dimensions.\\
In Section~2.1 we introduce classical LGT and make some general remarks on its
quantum deformation.\\
In Section~2.2 we collect facts from 3-manifold topology which are used
in the sequel.\\
In Section~2.3 we introduce the Reshetikhin-Turaev functor in the form
adopted for our purposes.\\
In Section~2.4 we define qQCD$_3$ functor and discuss the notion of gauge
invariance within our framework.\\
In Section~2.5 we introduce the qQCD$_3$ partition function in the case of
$\Uq(su(n))$ gauge group.\\
Section~2.6 is devoted to the root-of-unity case: $\Uq(sl(n,R))$,
$q^{\ell}=1$.\\
In Section~2.7 we prove that the weak-coupling partition function
introduced in the previous section is a topological invariant.\\
In Section~2.8 we discuss the case of bounded manifolds and shortly outline
the introduction of Wilson loop averages in our model.\\
Chapter~3 is devoted to the 2-dimensional case. Here we derive Verlinde's
formula and discuss a connection between our approach and the AGS algebra.\\
We conclude with a few general remarks.

\newsection{qQCD$_3$}
\subsection{Formulation of the problem}

To introduce lattice gauge theory, one needs a cell decomposition of a
manifold (in physicist usage, a lattice). A gauge field is a
map from a set of oriented edges to a compact group: 
$\ell\mapsto g_{\ell}\in G$. 
A change of an orientation corresponds to the conjugation: 
$g_{\ell}\to g_{\ell}^{\scs -1}$.

One attaches to every vertex a $G$-module (usually the regular
representation). Gauge transformations rotate bases of the modules
independently at each vertex.  The gauge field is interpreted as performing a
parallel transport between vertices, thus relating bases at adjacent ones.

If an oriented link, $k$, connects
vertices $v_2$ and $v_1$, the gauge transformation of the group element
$g_k$ is

\be
g_k\to h_{v_1}g_kh_{v_2}^{-1}
\label{gtrans}
\ee

A holonomy associated with a path $\{L\}$ in the lattice is an
ordered product of gauge field elements along $\{L\}$:

\be
h_{\scs L}=\prod_{k\in L} g_k^{\epsilon_k}
\label{arg}
\ee
where $\epsilon_k=+1$, if the $k$-th edge is directed along the path, and
$\epsilon_k=-1$, if their directions are opposite. 

Gauge invariant quantities are those taking values in the
set of conjugacy classes of $G$. A trace of the holonomy along
a closed loop in any representation of $G$ is an example of such an invariant.

The Boltzmann weights are functions of holonomies along boundaries, $\d f$, of
faces, $f$. One of the standard choices is the so-called group heat kernel

\be
W_\beta(h_{\d f})=\sum_R d_\R \chi_\R(h_{\d f}) e^{-\beta C_R},
\label{weight}
\ee

In eq. (\ref{weight}), 
$\sum_R$ is the sum over all finite dimensional irreps of a gauge group $G$; 
$\chi_\R(x)$ is the character of an irrep $R$; 
$d_R=\chi_\R(I)$ is its dimension; $C_R$ is a second Casimir eigenvalue; 
$\beta$ is a real parameter called a coupling constant.  
The construction makes sense for compact groups whose
unitary finite dimensional irreps span the regular 
representation.

The choice (\ref{weight}) ensures that $W_\beta(h_{\d f})$ becomes the
group $\delta$-function in the weak coupling limit, $\beta\to 0$:
$W_0(h_{\d f})=\delta(h_{\d f},I)$.
We shall call this limit {\em topological}.

The partition function is defined as the integral 
of the product of the Boltzmann weights over all faces:

\be
{\cal Z}_\beta=\int_G\prod_{\ell} dg_{\ell}\: 
\prod_f W_\beta(\prod_{k\in\partial f}g_k^{\epsilon_k})
\label{Z}
\ee
where $dg_{\ell}$ is the Haar measure on the group $G$, and the product
$\prod_{\ell}$ goes over all edges.

If the term ``q-deformed'' is to mean that gauge variables take values
in a quantum group, any presentation of the model should be reducible
to a form where the variables are represented in a standard fashion as
matrices of non-commutative elements.

The simplest and most famous example is $SL_q(2)$, which can be 
introduced as the set of matrices

\be
g=\left(\ba{cc} a&b\\ c&d \ea \right)
\label{qmatrix}
\ee
the entries of which obey the commutation relations
\be\ba{lll}
ba=qab & db=qbd & cb=bc \\
ca=qac & dc=qcd & da-ad=(q-q^{-1})bc\\
ad-q^{-1}bc&\!\! = 1&
\ea\label{comrel}
\ee

The relations (\ref{comrel}) imply the existence of the $R$-matrix

\be
R=\left(\ba{cccc}
q&0&0&0\\ 0&1&0&0\\ 0&q-q^{-1}&1&0\\ 0&0&0&q
\ea\right)\ee
and the $RTT=TTR$ equation

\be
(g\otimes 1)(1\otimes g)R=R(1\otimes g)(g\otimes 1)
\label{RTT=TTR}
\ee
The $R$-matrix obeys the quantum Yang-Baxter equation

\be
R_{12}R_{13}R_{23}=R_{23}R_{13}R_{12}
\label{YB}
\ee
Indices show at which positions in the tensor cube of representation
spaces, $\stackrel{1}{V}\otimes\stackrel{2}{V}\otimes\stackrel{3}{V}$,
acts the $R$-matrix.

$SL_q(2)$ has two real forms: $SU_q(2)$, for real $q$, 
and $SL_q(2,R)$, for $|q|=1$.

The matrices can be multiplied. If entries of both $g$ and $h$ obey 
\eq{comrel} and are mutually commutative, 
the entries of the product $gh$ obey (\ref{comrel}) as well. 
Therefore, matrices on different links of a lattice have to co-commute with 
one another in the tensor product.

The algebra of matrices (\ref{qmatrix}) naturally extends to the
quasi-triangular Hopf algebra $\Fq(SL(2))$ of quantized functions on
$SL(2)$. Owing to the famous duality, its basis is provided by the matrix
elements of finite-dimensional irreducible representations of the quantized
universal envelopping  (QUE) algebra $\Uq(sl(2))$.
Therefore, to construct qQCD, we have at hands co-multiplication,
$R$-matrix, antipode and Clebsch-Gordan coefficients (CGC). 

The q-deformation of \eq{Z} is roughly speaking a way to write it down in
terms of elements of the Hopf algebra. {\em A priori}, it is not
unique. The guiding principle here can be to identify any transformation of
the construction with some isometry of a base cell complex in a
self-consistent way. Then all algebraic manipulations become geometrically
meaningful. It is close in spirit to the Reshetikhin-Turaev
functor from the category of ribbon tangles to the modular Hopf algebras
\cite{ReshTur}. Our presentation of qQCD$_3$ is in many respects inspired
by their work. To describe it, we need to look at LGT from a bit more
general than usual point of view.

\subsection{Topological background}

For reader's convenience we collect in this section some
definitions which we shall use in the sequel.

A {\em $k$-cell} is a polyhedron homeomorphic to a $k$-dimensional ball.

A {\em cell complex} is a union of a finite number of cells such that
an intersection of any 2 $k$-cells is either empty or a finite
number of less dimensional cells.

A cell complex can be obtained starting with a finite set of points by
attaching subsequently cells of higher dimensions, any cell being
attached to a finite number of lower dimensional cells.

A union of all cells of dimension $\leq n$ is called an {\em
$n$-skeleton}.

A cell complex is a {\em manifold} if and only if the neighbourhood  of each
vertex is a spherical ball.

A complex is called {\em simplicial} if all cells are simplexes (\ie
points, links, triangles, tetrahedra, etc.).

Physicists usually mean by a {\em lattice} a cell complex such that an
intersection of any two $k$-cells either empty or consists of only one
entire less dimensional cell. We adopt this notion. Simplicial
complexes are lattices by definition.

A {\em dual} complex, $\wt{C}$, is constructed by putting into
correspondence its $k$-cells to cells of $C$ having complimentary
dimensions, $n-k$.

To introduce LGT, we need a presentation of a cell complex, \ie an
effective way to describe it unambiguously. In the classical case, one
needs to know only a 2-skeleton of a complex. 

From the topological point of view, the construction of LGT described at
the beginning of the previous section is reminiscent of the definition of
$H^1(C,G)$, the non-commutative first cohomology of $C$ with coefficients in
$G$. In the topological limit,
all holonomies along contractible loops vanish and gauge fields
obey the defining relations of $\pi_1(C)$. Therefore, being properly
normalized, the partition function ${\cal Z}_0$ counts the number of
conjugacy classes of injective homomorphisms from $\pi_1(C)$ into a gauge 
group $G$:

\be
{\cal Z}_0= |Hom(\pi_1(C),G)/G|
\label{fingr}
\ee

Of course, it makes sense only if $G$ is finite. If $G$ is a Lie
group, one speaks about a {\em moduli space of flat $G$-connections},
which is defined as a set of fields modulo gauge transformations:

\be
{\cal M}_G:= \{Hom(\pi_1(C),G)/G\}
\ee

It is easy to see that classical topological LGT is completely determined
by a homotopy type of a complex. The construction of qQCD$_3$ requires a more
precise presentation of a complex. 

It is known that any oriented 3-manifold
can be obtained by gluing up two 3-dimensional handlebodies along their
boundaries. This operation is the {\em Heegaard splitting}. The
minimal genus of the handlebodies is called the {\em Heegaard genus} of
the manifold.

We can obtain a Heegaard splitting for a given oriented manifold $M$
from its cellular decomposition, $C$, as follows. We take a tubular
neighbourhood, $H$, of the 1-skeleton of $C$. The complement of $H$ in
$M$, $\wt{H}=M\setminus H$, can be regarded as a tubular neighbourhood of the
1-skeleton of the dual complex $\wt{C}$. 

Every 1-cell $\si_i\in C$ determines a disk $D_i\subset H$ whose detachment
distroys a handle of $H$. The boundaries of the disks $\d D_i \subset \d H$
give a system of cycles on the boundary, $\d H$, of the handlebody $H$. We
shall call them the $\alpha$-cycles: $\alpha_i := \d D_i$. Dual 1-cells
$\wt{\si_j}\in\wt{C}$ determine analogously a system of
$\wt{\alpha}$-cycles on the boundary, $\d\wt{H}$, of $\wt{H}$. Images of
the $\wt{\alpha}$-cycles on $\d H$ produced by a gluing homomorphism $h$
are called the {\em characteristic curves} (or $\gamma$-cycles) of the
Heegaard diagram and define the manifold $M=H\bigcup_h\wt{H}$
unambiguously.

Let us fix a number of the disks $\{\wh{D}\}\subset\{D\}$ such that the
detachment of them makes the handlebody connected and simply-connected (\ie
$H\backslash \{\wh{D}\} \cong B^3$). We can put into correspondence a
generator $a_i$ of the fundamental group to each disk
$D_i\in\{\wh{D}\}$. Defining relations are read off in an obvious way from
a system of the characteristic curves $\{\gamma\}$. That is, if $\gamma_j$
intersects disks $D_{j_1},D_{j_2},\ldots,D_{j_k}$ subsequently, then the
corresponding relator is
$\Gamma_j=a_{j_1}^{\epsilon_{j_1}}a_{j_2}^{\epsilon_{j_2}}\ldots
a_{j_k}^{\epsilon_{j_k}}$, where $\epsilon_k=\pm1$ is the intersection
number depending on a mutual orientation of $\gamma_j$ and the $k$-th disk
at the intersection point. This set of relators is of course excessive. A
minimal set can be fixed by choosing a number of $\wt{\alpha}$-cycles which
span a disjoint collection of disks $\{\wh{\wt{D}}\}$ in the complementary
handlebody $\wt{H}=M\backslash H$ such that the detachment of all the
disks from the set $\{\wh{\wt{D}}\}$ makes $\wt{H}$ a 3-ball:
$\wt{H}\backslash\{\wh{\wt{D}}\} \cong B^3$.

One can deform a Heegaard diagram by any 2-dimensional isomorphism of a
boundary $\d H$ which extends to the whole handlebody $H$. A set of
generators for such isomorphisms is called in the literature the Suzuki
moves (see, \eg Refs.~\cite{Kohno,Crane} for an exposition accessible to a
physicist).

It can be shown that any class of
isotopic diffeomorphisms of a genus $g$ surface $M^2_g$ onto itself has a
representative which can be constructed as a composition of the Dehn twists,
$T^{\epsilon}_{\mu}$, where $\mu$ is one of the basic cycles on $M^2_g$ and
$\epsilon=\pm1$. One detaches from $M^2_g$ a thin neighborhood
$U_{\mu}\cong S^1\times [0,1]$ of a cycle $\mu$ and then attaches it back
after the full twist $U_{\mu}\to U_{\mu}$: $\varphi\times t \to
(\varphi+2\pi\epsilon t)\times t$ (where $t\in [0,1]$ and $\phi\in[0,2\pi]$
parametrizes $S^1$).

In the sequel, we shall only need the following fact: all the Suzuki moves are
combinations of Dehn twists on loops in $\d H$ which bound disks
$D\subset H$, except for the {\em handle slide} defined in the following
way. Imagine solid handles attached to a surface of a spherical ball. One
drags one end of a handle up, along and down  another handle. As a result of
this operation, an $\alpha$-cycle corresponding to
the second handle slides around an $\alpha$-cycle corresponding to the
first one. It can be described as
a multiplication of loops on $\d H$  defined in the standard fashion as
in the definition of the fundamental group $\pi_1(M_g^2)$

The same operations can be applied to $\wt{H}$ as well.

It is a classical result that any two Heegaard diagrams
representing the same manifold can be connected by a sequence of
the following operations:

\begin{enumerate}
\item Dehn twists on loops contractible in $H$ or $\wt{H}$. They do not
change a presentation of $\pi_1(C)$.

\item {\em Cycle slide,} which consists in the multiplication of a cycle by
another one: $\gamma_j\to \gamma_j\gamma_k$. It means that a relator
$\Gamma_j$ in a presentation of $\pi_1(C)$ is substituted by $\Gamma_j
\Gamma_k$. The same operation applied to the $\alpha$-cycles, $\alpha_j\to
\alpha_j\alpha_k$, corresponds to the change of generators of $\pi_1(C)$:
$a_j$ is substituted for $a_ja_k$.

\item {\em Stabilization}, which consists in adding a new handle to $H$ and
extending a gluing diffeomorphism by the identity on its boundary.
It means that one adds one characteristic curve and one 
$\alpha$-cycle to a Heegaard diagram or,
equivalently, a new generator $a_{g+1}$ along with the trivial relation,
$\Gamma_{g+1}=a_{g+1}=1$,  to a presentation of $\pi_1(C)$.
\end{enumerate}

It should be noted that an isotopy within a handlebody itself cannot
necessarily take place for its embeddings in $R^3$. The obvious obstruction
is that the characteristic curves can become linked in $R^3$.

We shall need the operation of the {\em connected sum} of two manifolds:
$M=M_1\# M_2$. One constructs $M$ by deleting spherical balls from $M_1$
and $M_2$ and then gluing the manifolds together along the
boundaries. Obviously, $M\#S^3\cong M$, which can be represented as the
attachment of a single 3-cell to the spherical boundary of the ball
obtained from $M$. This operation introduces an abelian semi-group
structure and any 3-fold invariant can be regarded as a representation of
this semi-group.

A manifold is called {\em simple}, if it cannot be represented as a
connected sum of two non-spherical manifolds. Any compact oriented
3-manifold possesses a unique expansion into a connected sum of simple
manifolds. 

By performing a Heegaard splitting, a manifold is constructed out of two
handlebodies joined by some homeomorphism of their boundaries. Such a
homeomorphism can be continued into small neighborhoods of the
boundaries. It means that the characteristic curves can submerge a bit into
the inside of $H$. If $n_{ij}=(\alpha_i,\gamma_j)$ is an intersection
number of two cycles on the boundary, then $\gamma_j$ will have after
the deformation the same number as a linking coefficient with
$\alpha_i$. The {\em linking coefficient} of two loops in $R^3$ is
equal, by definition, to the intersection number of the first with a
disk spanned by the second.

If $\alpha_i$ and $\gamma_j$ are linked, the
corresponding 1-cell, $\sigma^1_i$, enters the boundary of the
corresponding 2-cell, $\sigma^2_j$: $\sigma^1_i\in\d\sigma^2_j$. And
{\sl vice versa} for co-boundaries: $\sigma^2_j\in\delta\sigma^1_i$.
The boundary of a 2-cell defines a natural cyclic order of 1-cells
belonging to it. A peculiarity of the dimension 3 is that 2-cells
forming a co-boundary of a 1-cell are naturally ordered as well. It is a
cyclic order of dual 1-cells forming a boundary of a dual 2-cell.

\subsection{Reshetikhin-Turaev functor}

The quantized function (QF) algebra $\Fq(SL(n))$ is dual to the QUE algebra
$\Uq(sl(n))$, therefore the topological basis of $\Fq(SL(n))$ is given by
the matrix elements of irreducible representations of $\Uq(sl(n))$. For
example, the $2\times2$ matrix realization of $SL_q(2)$ given in
Eqs.~(\ref{qmatrix}) and (\ref{comrel}) exactly corresponds to the matrix
elements of the 2-dimensional irreducible representation of
$\Uq(sl(2))$. In this paper, we shall deal with real forms of
$\Uq(sl(n,C))$ with respect to some $\ast$-structures, whose existence is
always assumed.

The discussion in the previous section suggests that we can construct
qQCD$_3$ with help of the
Reshetikhin-Turaev functor from the category of colored ribbon tangles {\bf
ctang} to the category of representation rings of $\Uq(sl(n))$, 
{\bf rep$_{\Uq}$}.

The basic geometric object is a {\em tangle}, which was defined in
\rf{ReshTur} as ``a link of circles and segments in the 3-ball, where
it is assumed that ends of segments lie on the boundary of the ball''.
One puts into correspondence to every tangle a linear operator, $f$,
acting on a tensor product of modules associated with segments (which
have therefore to be oriented).

\be
f:\ V_{i_1}\otimes\ldots\otimes V_{i_n}
\to V_{j_1}\otimes\ldots\otimes V_{j_k}
\ee
or, graphically,
\newcounter{cnt}
\setlength{\unitlength}{1mm}
\be
f^{j_1\ldots j_k}_{i_1\ldots i_n}\ \cong
\raisebox{-2cm}{\bp(30,40)(0,-20) 
\thicklines
\put(5,-5){\framebox(20,10){$f$}}
\multiput(7,5)(4,0){5}{\vector(0,1){3}\line(0,1){5}}
\multiput(7,-10)(4,0){5}{\vector(0,1){3}\line(0,1){5}}
\put(6,-13){$\scr i_1$}
\put(22,-13){$\scr i_n$}
\put(6,12){$\scr j_1$}
\put(22,12){$\scr j_k$}
\ep}
\ee
where $j_1,\ldots,j_k$ and $i_1,\ldots,i_n$ are some indices
numerating modules.
The simplest example is the identity operator represented by a single
segment:

\be
\delta_{\alpha,\beta}\ \cong
\raisebox{-0.8cm}{\bp(10,20)
\setlength{\unitlength}{0.0125in}
\thicklines
\put(5,0){\shortstack{$\alpha$\\ \rule{0cm}{1cm}
\\ \rule{0cm}{0.4cm}$\beta$}}
\drawline(5,10)(5,50)
\drawline(7.000,26.000)(5.000,34.000)(3.000,26.000)
\ep}
\ee

All modules considered in this paper are assumed to be irreducible. We
shall draw linear operators acting on them as small boxes (coupons) with
labels inside. The elementary building blocks are the matrix elements

\be
D^i_{\alpha,\beta}(a)\ \cong
\raisebox{-0.5in}{
\setlength{\unitlength}{0.0125in}
\begin{picture}(23,85)(0,-10)
\thicklines
\drawline(20,45)(20,25)(0,25)
	(0,45)(20,45)
\drawline(10,25)(10,15)(10,15)
\drawline(10,70)(10,60)
\drawline(10,45)(10,60)
\drawline(12.000,52.000)(10.000,60.000)(8.000,52.000)
\put(10,35){\makebox(0,0){$a$}}
\drawline(10,0)(10,15)
\drawline(12.000,7.000)(10.000,15.000)(8.000,7.000)
\put(14,-5){\makebox(0,0){$i,\beta$}}
\put(14,77){\makebox(0,0){$\alpha$}}
\end{picture}}
\ee
where $a$ is an element of $\Uq$. The arrows show a direction of 
the action of an operator. We use the Greek letters to numerate basis
vectors of irreducible modules and the Latin ones, to numerate the modules. 
They will often be omited.

$\Uq$ possesses several $\ast$-structures. We shall draw conjugate objects as

\be
D^i_{\alpha,\beta}(a^*)\ \cong
\raisebox{-0.5in}{
\setlength{\unitlength}{0.0125in}
\begin{picture}(23,85)(0,-10)
\thicklines
\drawline(20,45)(20,25)(0,25)
	(0,45)(20,45)
\drawline(10,25)(10,15)(10,15)
\drawline(10,70)(10,60)
\drawline(10,45)(10,60)
\drawline(12.000,60.000)(10.000,52.000)(8.000,60.000)
\put(10,35){\makebox(0,0){$a$}}
\drawline(10,0)(10,15)
\drawline(12.000,15.000)(10.000,7.000)(8.000,15.000)
\put(12,-6){\makebox(0,0){$i,\alpha$}}
\put(12,77){\makebox(0,0){$\beta$}}
\end{picture}}
\mbox{\Large =}
\raisebox{-0.5in}{
\setlength{\unitlength}{0.0125in}
\begin{picture}(23,85)(0,-10)
\thicklines
\drawline(20,45)(20,25)(0,25)
	(0,45)(20,45)
\drawline(10,25)(10,15)(10,15)
\drawline(10,70)(10,60)
\drawline(10,45)(10,60)
\drawline(12.000,52.000)(10.000,60.000)(8.000,52.000)
\put(10,35){\makebox(0,0){$a$}}
\drawline(10,0)(10,15)
\drawline(12.000,7.000)(10.000,15.000)(8.000,7.000)
\put(12,-6){\makebox(0,0){$i,\overline{\alpha}$}}
\put(12,77){\makebox(0,0){$\overline{\beta}$}}
\end{picture}}
\ee
The last equality takes place for a real form of $\Uq$, where the
$\ast$-structure matches basis vectors of a module:
$\alpha\to\overline{\alpha}$.  

The operators form an {\em algebra} $\al$. We can translate this
property in pictures as

\be
\bigg\{a,b\in\al\to ab\in\al\bigg\}\cong 
\bigg\{
\setlength{\unitlength}{0.0125in}
\raisebox{-0.85in}{
\setlength{\unitlength}{0.0125in}
\begin{picture}(109,139)(0,-10)
\thicklines
\path(20,95)(20,75)(0,75)
	(0,95)(20,95)
\path(10,50)(10,65)
\path(12.000,57.000)(10.000,65.000)(8.000,57.000)
\path(10,75)(10,65)(10,65)
\path(10,95)(10,110)
\path(12.000,102.000)(10.000,110.000)(8.000,102.000)
\path(20,50)(20,30)(0,30)
	(0,50)(20,50)
\path(10,30)(10,20)(10,20)
\path(10,5)(10,20)
\path(12.000,12.000)(10.000,20.000)(8.000,12.000)
\path(100,70)(100,50)(80,50)
	(80,70)(100,70)
\path(90,50)(90,40)(90,40)
\path(10,120)(10,110)
\path(90,25)(90,40)
\path(92.000,32.000)(90.000,40.000)(88.000,32.000)
\put(90,60){\makebox(0,0){$ab$}}
\path(90,70)(90,85)
\path(92.000,77.000)(90.000,85.000)(88.000,77.000)
\path(90,95)(90,85)(90,85)
\path(40,65)(60,65)
\path(40,60)(60,60)
\put(10,85){\makebox(0,0){$a$}}
\put(10,40){\makebox(0,0){$b$}}
\end{picture}}
\bigg\}
\ee
\be
\bigg\{\exists1\in\al: a1=1a=a,\ \forall 
a\in\al\bigg\} \cong \bigg\{
\setlength{\unitlength}{0.0125in}
\raisebox{-0.55in}{
\begin{picture}(183,90)(0,-10)
\thicklines
\path(90,25)(90,15)(90,15)
\path(90,0)(90,15)
\path(92.000,7.000)(90.000,15.000)(88.000,7.000)
\path(90,45)(90,60)
\path(92.000,52.000)(90.000,60.000)(88.000,52.000)
\path(90,70)(90,60)(90,60)
\path(40,40)(60,40)
\path(40,35)(60,35)
\path(20,45)(20,25)(0,25)
	(0,45)(20,45)
\path(10,0)(10,15)
\path(12.000,7.000)(10.000,15.000)(8.000,7.000)
\path(10,25)(10,15)(10,15)
\path(10,45)(10,60)
\path(12.000,52.000)(10.000,60.000)(8.000,52.000)
\path(10,70)(10,60)(10,60)
\path(100,45)(100,25)(80,25)
	(80,45)(100,45)
\path(120,35)(140,35)
\put(10,35){\makebox(0,0){$1a$}}
\path(120,40)(140,40)
\path(100,45)(100,25)(80,25)
	(80,45)(100,45)
\path(170,5)(170,20)
\path(172.000,12.000)(170.000,20.000)(168.000,12.000)
\path(90,0)(90,15)
\path(92.000,7.000)(90.000,15.000)(88.000,7.000)
\path(170,75)(170,65)(170,65)
\path(170,50)(170,65)
\path(172.000,57.000)(170.000,65.000)(168.000,57.000)
\path(170,30)(170,20)(170,20)
\path(180,50)(180,30)(160,30)
	(160,50)(180,50)
\put(90,35){\makebox(0,0){$a1$}}
\put(170,40){\makebox(0,0){$a$}}
\end{picture}}
\bigg\}
\ee

$\al$ is a ring, \ie an Abelian group under some $+$ operation, which
we shall understand as a formal sum of pictures with the natural
definition of the multiplication by an integer number.

To have a {\em bi-algebra} structure on $\al$, we need a {\em
co-multiplication} $\Delta:\ V\to V\otimes V$ and a {\em co-unit}
$\eps$. We introduce $\Delta$ as

\be
\Delta(a) \cong 
\raisebox{-0.7in}{
\setlength{\unitlength}{0.0125in}
\begin{picture}(130,115)(25,-10)
\thicklines
\path(35,15)(35,40)
\path(37.000,22.000)(35.000,30.000)(33.000,22.000)
\path(35,60)(35,85)
\path(37.000,67.000)(35.000,75.000)(33.000,67.000)
\path(45,15)(45,40)
\path(47.000,22.000)(45.000,30.000)(43.000,22.000)
\path(45,60)(45,85)
\path(47.000,67.000)(45.000,75.000)(43.000,67.000)
\path(55,60)(55,40)(25,40)
	(25,60)(55,60)
\put(40,50){\makebox(0,0){$\Delta(a)$}}
\path(65,52)(80,52)
\path(65,48)(80,48)
\path(135,60)(135,40)(115,40)
	(115,60)(135,60)
\put(95,45){\makebox(0,0){$\displaystyle \sum_i$}}
\put(135,80){\makebox(0,0){$i$}}
\put(125,50){\makebox(0,0){$a$}}
\path(125,40)(125,30)(125,30)
\path(125,15)(125,30)
\path(127.000,22.000)(125.000,30.000)(123.000,22.000)
\path(125,60)(125,75)
\path(127.000,67.000)(125.000,75.000)(123.000,67.000)
\path(125,85)(125,75)(125,75)
\path(125,85)(110,100)
\path(125,85)(140,100)
\path(125,15)(140,0)
\path(125,15)(110,0)
\end{picture}
}\mbox{\Huge =}
\raisebox{-0.7in}{
\setlength{\unitlength}{0.0125in}
\begin{picture}(65,115)(0,-10)
\thicklines
\path(10,15)(10,30)
\path(12.000,22.000)(10.000,30.000)(8.000,22.000)
\path(10,40)(10,30)(10,30)
\path(10,60)(10,75)
\path(12.000,67.000)(10.000,75.000)(8.000,67.000)
\path(10,85)(10,75)(10,75)
\path(40,15)(40,30)
\path(42.000,22.000)(40.000,30.000)(38.000,22.000)
\path(40,40)(40,30)(40,30)
\path(40,60)(40,75)
\path(42.000,67.000)(40.000,75.000)(38.000,67.000)
\path(40,85)(40,75)(40,75)
\path(50,60)(50,40)(30,40)
	(30,60)(50,60)
\path(20,60)(20,40)(0,40)
	(0,60)(20,60)
\put(10,50){\makebox(0,0){$a$}}
\put(40,50){\makebox(0,0){$a$}}
\end{picture}
}
\label{co-mult}
\ee
In general, the last equality is simply a convenient pictorial representation  
and has to be given a precise meaning in every particular case.

In \eq{co-mult}, the 3-valent vertices,

\be
\raisebox{-1cm}{\bp(25,30)
\thicklines
\put(10,10){\line(1,-1){7}}
\put(10,10){\line(-1,-1){7}}
\put(10,10){\line(0,1){10}}
\put(10,22){$j_3,\alpha_3$}
\put(0,0){$j_1,\alpha_1$}\put(17,0){$j_2,\alpha_2$}
\ep}
\cong C^{j_3\alpha_3}_{j_1\alpha_1;j_2\alpha_2}
\hspace{4pc}
\raisebox{-1cm}{\bp(25,30)
\thicklines
\put(10,10){\line(1,1){7}}
\put(10,10){\line(-1,1){7}}
\put(10,10){\line(0,-1){7}}
\put(7,0){$j_3,\alpha_3$}
\put(0,20){$j_1,\alpha_1$}\put(18,20){$j_2,\alpha_2$}
\ep}
\cong \overline{C}^{j_1\alpha_1;j_2\alpha_2}_{j_3\alpha_3}
\ee
are the quantum Clebsch-Gordan coefficients 
$\displaystyle e^{j_1}_{\alpha_1}\otimes e^{j_2}_{\alpha_2}=
\sum_{j_3,\alpha_3}C^{j_3\alpha_3}_{j_1\alpha_1;j_2\alpha_2}
e^{j_3}_{\alpha_3}$ ($e^{j}_{\alpha}$ is a basis of $V_j$).
They obey the properties

\be
\sum_i\!\!\!\!\!\!\!\!\raisebox{-0.75in}{
\setlength{\unitlength}{0.0125in}
\begin{picture}(243,110)(0,-10)
\thicklines
\path(32.000,47.000)(30.000,55.000)(28.000,47.000)
\path(30,55)(30,35)(15,15)
\path(30,35)(45,15)
\path(30,55)(30,70)(45,90)
\path(30,70)(15,90)
\path(90,15)(90,55)
\path(92.000,47.000)(90.000,55.000)(88.000,47.000)
\path(110,15)(110,55)
\path(112.000,47.000)(110.000,55.000)(108.000,47.000)
\path(90,55)(90,90)
\path(110,55)(110,90)
\path(170,15)(170,30)
\path(172.000,22.000)(170.000,30.000)(168.000,22.000)
\path(170,70)(170,85)
\path(172.000,77.000)(170.000,85.000)(168.000,77.000)
\path(170,40)(170,30)
\put(170,55){\ellipse{20}{30}}
\path(170,95)(170,85)
\put(125,55){\makebox(0,0){\Huge ;}}
\path(225,15)(225,60)
\path(227.000,52.000)(225.000,60.000)(223.000,52.000)
\path(225,60)(225,95)
\put(65,55){\makebox(0,0){\Huge =}}
\put(110,0){\makebox(0,0){$j_2$}}
\put(90,0){\makebox(0,0){$j_1$}}
\put(50,0){\makebox(0,0){$j_2$}}
\put(10,0){\makebox(0,0){$j_1$}}
\put(40,60){\makebox(0,0){$i$}}
\put(195,55){\makebox(0,0){\Huge =}}
\put(170,0){\makebox(0,0){$j$}}
\put(225,0){\makebox(0,0){$j$}}
\end{picture}
}
\ee
which simply means that they are elements of a unitary matrix
connecting bases in $V$ and $V\otimes V$:

\[
\sum_{i,\beta}C^{i\beta}_{j_1\alpha_1;j_2\alpha_2}
\overline{C}^{j'_1\alpha'_1;j'_2\alpha'_2}_{i\beta}=
\delta_{j_1,j'_1}\delta_{j_2,j'_2}
\delta_{\alpha_1,\alpha'_1}\delta_{\alpha_2,\alpha'_2}
\]

We can check the properties of the co-multiplication graphically

\be
\bigg\{\Delta(ab)=\Delta(a)\Delta(b)\bigg\}\cong
\bigg\{\raisebox{-1.0in}{
\setlength{\unitlength}{0.0125in}
\begin{picture}(170,145)(0,-10)
\thicklines
\path(25,80)(25,60)(5,60)
	(5,80)(25,80)
\path(15,35)(15,50)
\path(17.000,42.000)(15.000,50.000)(13.000,42.000)
\path(15,60)(15,50)(15,50)
\path(15,80)(15,95)
\path(17.000,87.000)(15.000,95.000)(13.000,87.000)
\path(15,105)(15,95)(15,95)
\path(15,105)(30,120)
\path(15,35)(30,20)
\path(15,35)(0,20)
\path(15,105)(0,120)
\path(85,95)(85,75)(65,75)
	(65,95)(85,95)
\path(85,60)(85,40)(65,40)
	(65,60)(85,60)
\path(75,40)(75,30)(75,30)
\path(15,35)(0,20)
\path(75,110)(90,125)
\path(75,60)(75,75)
\path(105,70)(120,70)
\path(145,15)(130,0)
\path(145,15)(160,0)
\path(75,110)(60,125)
\put(145,65){\ellipse{20}{20}}
\path(145,115)(130,130)
\put(145,35){\makebox(0,0){$b$}}
\path(145,115)(160,130)
\path(155,45)(155,25)(135,25)
	(135,45)(155,45)
\path(145,55)(145,45)
\path(145,75)(145,85)
\path(155,105)(155,85)(135,85)
	(135,105)(155,105)
\path(145,115)(145,105)
\path(145,25)(145,15)
\path(75,20)(60,5)
\path(75,20)(90,5)
\path(75,20)(75,30)
\path(77.000,22.000)(75.000,30.000)(73.000,22.000)
\path(75,95)(75,105)
\path(77.000,97.000)(75.000,105.000)(73.000,97.000)
\path(75,105)(75,110)
\path(35,65)(50,65)
\path(35,70)(50,70)
\path(105,65)(120,65)
\put(15,70){\makebox(0,0){$ab$}}
\put(75,85){\makebox(0,0){$a$}}
\put(75,50){\makebox(0,0){$b$}}
\put(145,95){\makebox(0,0){$a$}}
\end{picture}
}
\bigg\}
\ee
and

\be
\bigg\{\Delta(1)=1\otimes1\bigg\}\cong\bigg\{
\raisebox{-0.6in}{
\setlength{\unitlength}{0.0125in}
\begin{picture}(100,90)(0,-10)
\thicklines
\path(15,20)(30,0)
\path(15,40)(15,55)(30,75)
\path(15,55)(0,75)
\path(17.000,32.000)(15.000,40.000)(13.000,32.000)
\path(15,40)(15,20)(0,0)
\path(75,0)(75,40)
\path(77.000,32.000)(75.000,40.000)(73.000,32.000)
\put(50,40){\makebox(0,0){\Huge =}}
\path(95,0)(95,40)
\path(97.000,32.000)(95.000,40.000)(93.000,32.000)
\path(75,40)(75,75)
\path(95,40)(95,75)
\end{picture}
}\ \bigg\}
\ee

In these formulas, the sum over intermediate states is assumed.
In what follows, we shall often omit the sum sign in pictures.

Thus, the co-associativity is coded in the properties of the
Clebsch-Gordan coefficients.

The co-unit is a homomorphism to an abelian group associated with a
field over which $\al$ is defined. 

\be
\eps(ab)=\eps(a)\eps(b)\hspace{1pc}\eps(1)=1
\ee
We shall connect the co-unit
with a projection on the trivial representation of a quantum group. In
other words, with the group integration.

To have the {\em Hopf algebra} structure on $\al$, we introduce an
{\em antipode} map: $S:\ \al\to\al$:

\be
S\Big(
\raisebox{-0.5in}{
\setlength{\unitlength}{0.0125in}
\begin{picture}(24,85)(6,-10)
\thicklines
\path(15,0)(15,15)
\path(17.000,7.000)(15.000,15.000)(13.000,7.000)
\path(15,25)(15,15)(15,15)
\path(15,45)(15,60)
\path(17.000,52.000)(15.000,60.000)(13.000,52.000)
\path(15,70)(15,60)(15,60)
\path(25,45)(25,25)(5,25)
	(5,45)(25,45)
\put(15,35){\makebox(0,0){$a$}}
\end{picture}}\Big)
\mbox{\Huge =}
\raisebox{-0.5in}{
\setlength{\unitlength}{0.0125in}
\begin{picture}(50,85)(60,-10)
\thicklines
\path(95,45)(95,25)(75,25)
	(75,45)(95,45)
\put(85,35){\makebox(0,0){$a$}}
\path(105,5)(105,25)
\path(107.000,17.000)(105.000,25.000)(103.000,17.000)
\path(65,30)(65,45)
\path(67.000,37.000)(65.000,45.000)(63.000,37.000)
\path(65,45)(65,65)
\path(85,45)	(84.565,48.675)
	(84.419,51.411)
	(85.000,55.000)

\path(85,55)	(86.405,58.049)
	(88.485,61.322)
	(91.322,63.933)
	(95.000,65.000)

\path(95,65)	(98.678,63.933)
	(101.515,61.321)
	(103.595,58.049)
	(105.000,55.000)

\path(105,55)	(105.435,52.707)
	(105.387,49.920)
	(105.000,45.000)

\path(105,45)	(105.000,42.062)
	(105.000,38.062)
	(105.000,35.517)
	(105.000,32.531)
	(105.000,29.045)
	(105.000,25.000)

\path(85,25)	(85.435,21.325)
	(85.581,18.589)
	(85.000,15.000)

\path(85,15)	(83.595,11.951)
	(81.515,8.679)
	(78.678,6.067)
	(75.000,5.000)

\path(75,5)	(71.322,6.067)
	(68.485,8.679)
	(66.405,11.951)
	(65.000,15.000)

\path(65,15)	(64.347,17.338)
	(64.129,20.383)
	(64.347,24.487)
	(64.619,27.045)
	(65.000,30.000)

\end{picture}} \mbox{\Huge =}
\raisebox{-0.5in}{
\setlength{\unitlength}{0.0125in}
\begin{picture}(30,85)(0,-10)
\thicklines
\path(15,0)(15,15)
\path(17.000,7.000)(15.000,15.000)(13.000,7.000)
\path(15,25)(15,15)(15,15)
\path(15,45)(15,60)
\path(17.000,52.000)(15.000,60.000)(13.000,52.000)
\path(15,70)(15,60)(15,60)
\path(25,45)(25,25)(5,25)
	(5,45)(25,45)
\put(15,35){\makebox(0,0){$a^{\ast}$}}
\end{picture}}
\ee
obeying

\be
\cdot(S\otimes id)\circ\Delta = \cdot(id\otimes S)\circ\Delta =
1\circ\eps 
\ee
which looks graphically as

\be
\cdot(S\otimes id)\circ
\raisebox{-0.85in}{
\setlength{\unitlength}{0.0125in}
\begin{picture}(180,135)(10,-10)
\thicklines
\path(20,70)(20,90)
\path(22.000,82.000)(20.000,90.000)(18.000,82.000)
\path(0,120)(20,100)(40,120)
\path(20,100)(20,90)
\path(20,20)(20,40)
\path(22.000,32.000)(20.000,40.000)(18.000,32.000)
\path(0,0)(20,20)(40,0)
\path(20,50)(20,40)
\path(120,50)(120,70)(100,70)
	(100,50)(120,50)
\path(110,70)(110,85)
\path(112.000,77.000)(110.000,85.000)(108.000,77.000)
\path(110,50)(110,40)
\path(110,20)(110,40)
\path(112.000,32.000)(110.000,40.000)(108.000,32.000)
\path(30,50)(30,70)(10,70)
	(10,50)(30,50)
\path(110,20)(125,5)
\put(145,60){\makebox(0,0){\Huge =}}
\path(175,20)(175,65)
\path(177.000,57.000)(175.000,65.000)(173.000,57.000)
\path(175,65)(175,100)
\path(110,95)(110,85)
\path(75,60)(75,65)
\path(77.000,57.000)(75.000,65.000)(73.000,57.000)
\path(110,20)	(106.235,16.235)
	(103.469,13.469)
	(100.000,10.000)

\path(100,10)	(97.734,7.347)
	(95.000,5.000)

\path(95,5)	(92.737,4.419)
	(90.000,4.226)
	(87.263,4.419)
	(85.000,5.000)

\path(85,5)	(82.156,6.683)
	(79.226,9.226)
	(76.683,12.156)
	(75.000,15.000)

\path(75,15)	(74.565,17.293)
	(74.613,20.080)
	(75.000,25.000)

\path(75,25)	(75.000,27.203)
	(75.000,30.000)
	(75.000,32.797)
	(75.000,35.000)

\path(75,35)	(75.000,37.476)
	(75.000,40.508)
	(75.000,43.911)
	(75.000,47.500)
	(75.000,51.089)
	(75.000,54.492)
	(75.000,57.524)
	(75.000,60.000)

\path(75,60)	(75.000,64.407)
	(75.000,67.164)
	(75.000,70.407)
	(75.000,74.224)
	(75.000,78.704)
	(75.000,81.219)
	(75.000,83.933)
	(75.000,86.856)
	(75.000,90.000)

\path(110,95)	(106.734,98.266)
	(105.000,100.000)

\path(105,100)	(102.347,102.266)
	(100.000,105.000)

\path(100,105)	(99.002,109.945)
	(99.167,112.675)
	(100.000,115.000)

\path(100,115)	(104.485,118.605)
	(107.345,119.633)
	(110.000,120.000)

\path(110,120)	(112.655,119.633)
	(115.515,118.605)
	(120.000,115.000)

\path(120,115)	(120.833,112.675)
	(120.998,109.945)
	(120.000,105.000)

\path(120,105)	(117.653,102.266)
	(115.000,100.000)

\path(115,100)	(113.266,98.266)
	(110.000,95.000)

\put(20,60){\makebox(0,0){$a$}}
\put(55,60){\makebox(0,0){\Huge =}}
\put(110,60){\makebox(0,0){$a$}}
\end{picture}
}\eps(a)
\ee
where $a$ is an arbitrary element of $\al$ and $\circ$ means a composition
of operations. This property shows that the antipode can serve as a
$q$-analog of the inverse. However, in general, $S^2\neq1$. 

The maps $V\otimes V \to C$ and $C \to V\otimes V$ are constructed with
help of CGC:

\be
\raisebox{-0.7in}{
\setlength{\unitlength}{0.0125in}
\begin{picture}(250,100)(-10,-10)
\thicklines
\path(15,25)(15,40)
\path(17.000,32.000)(15.000,40.000)(13.000,32.000)
\path(60,20)(65,30)
\path(63.211,21.950)(65.000,30.000)(59.633,23.739)
\path(90,20)(85,30)
\path(90.367,23.739)(85.000,30.000)(86.789,21.950)
\path(65,30)(75,45)(85,30)
\path(75,45)(75,65)
\path(130,40)(130,45)
\path(132.000,37.000)(130.000,45.000)(128.000,37.000)
\path(155,40)(155,45)
\path(157.000,37.000)(155.000,45.000)(153.000,37.000)
\path(130,45)(130,55)
\path(155,45)(155,55)
\path(-10,25)(-10,40)
\path(-8.000,32.000)(-10.000,40.000)(-12.000,32.000)
\path(220,15)(220,35)(230,50)
\path(227.226,42.234)(230.000,50.000)(223.898,44.453)
\path(230,50)(240,65)
\path(210,50)(200,65)
\path(220,35)(210,50)
\path(216.102,44.453)(210.000,50.000)(212.774,42.234)
\path(-10,40)	(-10.157,43.243)
	(-10.000,45.000)

\path(-10,45)	(-10.630,47.420)
	(-9.628,50.306)
	(-5.000,55.000)

\path(-5,55)	(-2.456,56.375)
	(2.500,56.833)
	(6.544,56.375)
	(10.000,55.000)

\path(10,55)	(13.372,50.306)
	(14.370,47.420)
	(15.000,45.000)

\path(15,45)	(15.157,43.243)
	(15.000,40.000)

\path(130,40)	(129.843,36.757)
	(130.000,35.000)

\path(130,35)	(130.630,32.580)
	(131.628,29.694)
	(135.000,25.000)

\path(135,25)	(138.456,23.625)
	(142.500,23.167)
	(146.544,23.625)
	(150.000,25.000)

\path(150,25)	(153.372,29.694)
	(154.370,32.580)
	(155.000,35.000)

\path(155,35)	(155.157,36.757)
	(155.000,40.000)

\put(40,40){\makebox(0,0){$:=\sqrt{d_j}$}}
\put(100,40){\makebox(0,0){\Huge ;}}
\put(180,40){\makebox(0,0){$:=\sqrt{d_j}$}}
\put(75,75){\makebox(0,0){$\scr 0$}}
\put(220,5){\makebox(0,0){$\scr 0$}}
\put(55,10){\makebox(0,0){$\scr j$}}
\put(95,10){\makebox(0,0){$\scr j$}}
\put(-10,15){\makebox(0,0){$\scr j$}}
\put(15,15){\makebox(0,0){$\scr j$}}
\put(195,75){\makebox(0,0){$\scr j$}}
\put(245,75){\makebox(0,0){$\scr j$}}
\put(130,65){\makebox(0,0){$\scr j$}}
\put(155,65){\makebox(0,0){$\scr j$}}
\end{picture}}
\ee
where $d_j$ is the quantum dimension of the module $V_j$. These objects
become the ordinary $\delta$-functions in the $q\to1$ limit.

The self-consistency requires that

\be
\bigg\{S(ab)=S(a)S(b)\bigg\}\cong
\Bigg\{
\raisebox{-0.65in}{
\setlength{\unitlength}{0.0125in}
\begin{picture}(246,102)(0,-10)
\thicklines
\path(1,40)(1,55)
\path(3.000,47.000)(1.000,55.000)(-1.000,47.000)
\path(31,55)(31,35)(11,35)
	(11,55)(31,55)
\path(41,15)(41,45)
\path(43.000,37.000)(41.000,45.000)(39.000,37.000)
\path(51,40)(66,40)
\path(51,45)(66,45)
\path(106,40)(106,20)(86,20)
	(86,40)(106,40)
\path(106,70)(106,50)(86,50)
	(86,70)(106,70)
\path(76,25)(76,40)
\path(78.000,32.000)(76.000,40.000)(74.000,32.000)
\path(76,40)(76,75)
\path(96,40)(96,50)
\path(116,0)(116,50)
\path(118.000,42.000)(116.000,50.000)(114.000,42.000)
\path(126,45)(141,45)
\path(126,40)(141,40)
\path(156,55)(156,75)
\path(156,40)(156,55)
\path(158.000,47.000)(156.000,55.000)(154.000,47.000)
\path(1,55)(1,75)
\path(186,55)(186,35)(166,35)
	(166,55)(186,55)
\put(96,60){\makebox(0,0){$b$}}
\path(226,55)(226,35)(206,35)
	(206,55)(226,55)
\path(236,10)(236,35)
\path(238.000,27.000)(236.000,35.000)(234.000,27.000)
\path(21,35)	(21.435,31.325)
	(21.581,28.589)
	(21.000,25.000)

\path(21,25)	(19.595,21.951)
	(17.515,18.679)
	(14.678,16.067)
	(11.000,15.000)

\path(11,15)	(7.322,16.067)
	(4.485,18.679)
	(2.405,21.951)
	(1.000,25.000)

\path(1,25)	(0.347,27.338)
	(0.129,30.383)
	(0.347,34.487)
	(0.619,37.045)
	(1.000,40.000)

\spline(21,55)
(21,65)(31,75)
	(41,65)(41,45)
\path(96,20)	(96.435,16.325)
	(96.581,13.589)
	(96.000,10.000)

\path(96,10)	(94.595,6.951)
	(92.515,3.679)
	(89.678,1.067)
	(86.000,0.000)

\path(86,0)	(82.322,1.067)
	(79.485,3.679)
	(77.405,6.951)
	(76.000,10.000)

\path(76,10)	(75.347,12.338)
	(75.129,15.383)
	(75.347,19.487)
	(75.619,22.045)
	(76.000,25.000)

\spline(96,70)
(96,80)(106,90)
	(116,80)(116,50)
\path(176,35)	(176.435,31.325)
	(176.581,28.589)
	(176.000,25.000)

\path(176,25)	(174.595,21.951)
	(172.515,18.679)
	(169.678,16.067)
	(166.000,15.000)

\path(166,15)	(162.322,16.067)
	(159.485,18.679)
	(157.405,21.951)
	(156.000,25.000)

\path(156,25)	(155.347,27.338)
	(155.129,30.383)
	(155.347,34.487)
	(155.619,37.045)
	(156.000,40.000)

\spline(216,55)
(216,65)(226,75)
	(236,65)(236,35)
\path(216,35)	(216.435,31.325)
	(216.581,28.589)
	(216.000,25.000)

\path(216,25)	(214.595,21.951)
	(212.515,18.679)
	(209.678,16.067)
	(206.000,15.000)

\path(206,15)	(202.322,16.067)
	(199.485,18.679)
	(197.405,21.951)
	(196.000,25.000)

\path(196,25)	(195.129,28.118)
	(194.839,32.178)
	(194.912,34.708)
	(195.129,37.649)
	(195.492,41.060)
	(196.000,45.000)

\path(196.892,36.802)(196.000,45.000)(192.929,37.347)
\path(176,55)	(175.565,58.675)
	(175.419,61.411)
	(176.000,65.000)

\path(176,65)	(177.405,68.049)
	(179.485,71.322)
	(182.322,73.933)
	(186.000,75.000)

\path(186,75)	(189.678,73.933)
	(192.515,71.321)
	(194.595,68.049)
	(196.000,65.000)

\path(196,65)	(196.871,61.882)
	(197.161,57.822)
	(197.088,55.292)
	(196.871,52.351)
	(196.508,48.940)
	(196.000,45.000)

\put(96,30){\makebox(0,0){$a$}}
\put(176,45){\makebox(0,0){$a$}}
\put(216,45){\makebox(0,0){$b$}}
\put(21,45){\makebox(0,0){$ab$}}
\end{picture}
}
\Bigg\}
\ee
For this property to hold, it is important that 

\be
\raisebox{-0.7in}{
\setlength{\unitlength}{0.0125in}
\begin{picture}(240,115)(0,-10)
\thicklines
\path(10,45)(10,55)
\path(20,75)(20,55)(0,55)
	(0,75)(20,75)
\path(10,75)(10,90)
\path(12.000,82.000)(10.000,90.000)(8.000,82.000)
\path(10,0)(10,15)
\path(12.000,7.000)(10.000,15.000)(8.000,7.000)
\path(10,15)(10,25)
\path(10,90)(10,100)
\path(145,55)(160,55)
\path(145,60)(160,60)
\path(200,65)(200,45)(180,45)
	(180,65)(200,65)
\path(240,65)(240,45)(220,45)
	(220,65)(240,65)
\path(190,45)(190,35)
\path(190,20)(190,35)
\path(192.000,27.000)(190.000,35.000)(188.000,27.000)
\path(230,65)(230,80)
\path(232.000,72.000)(230.000,80.000)(228.000,72.000)
\path(230,80)(230,90)
\path(20,45)(20,25)(0,25)
	(0,45)(20,45)
\path(40,50)(55,50)
\put(120,55){\makebox(0,0){$b$}}
\path(40,60)(55,60)
\path(80,80)(80,90)
\path(80,65)(80,80)
\path(82.000,72.000)(80.000,80.000)(78.000,72.000)
\path(90,65)(90,45)(70,45)
	(70,65)(90,65)
\path(130,65)(130,45)(110,45)
	(110,65)(130,65)
\path(120,20)(120,35)
\path(122.000,27.000)(120.000,35.000)(118.000,27.000)
\path(120,45)(120,35)
\path(190,65)	(189.565,68.675)
	(189.419,71.411)
	(190.000,75.000)

\path(190,75)	(191.405,78.049)
	(193.485,81.322)
	(196.322,83.933)
	(200.000,85.000)

\path(200,85)	(203.678,83.933)
	(206.515,81.321)
	(208.595,78.049)
	(210.000,75.000)

\path(210,75)	(210.871,70.413)
	(210.907,67.683)
	(210.774,64.840)
	(210.544,62.016)
	(210.290,59.346)
	(210.000,55.000)

\path(210,55)	(209.710,50.654)
	(209.456,47.984)
	(209.226,45.160)
	(209.093,42.317)
	(209.129,39.587)
	(210.000,35.000)

\path(210,35)	(211.405,31.951)
	(213.485,28.679)
	(216.322,26.067)
	(220.000,25.000)

\path(220,25)	(223.678,26.067)
	(226.515,28.679)
	(228.595,31.951)
	(230.000,35.000)

\path(230,35)	(230.581,38.589)
	(230.435,41.325)
	(230.000,45.000)

\path(80,45)	(79.565,41.325)
	(79.419,38.589)
	(80.000,35.000)

\path(80,35)	(81.405,31.951)
	(83.485,28.679)
	(86.322,26.067)
	(90.000,25.000)

\path(90,25)	(93.678,26.067)
	(96.515,28.679)
	(98.595,31.951)
	(100.000,35.000)

\path(100,35)	(100.871,39.587)
	(100.907,42.317)
	(100.774,45.160)
	(100.544,47.984)
	(100.290,50.654)
	(100.000,55.000)

\path(100,55)	(99.710,59.346)
	(99.456,62.016)
	(99.226,64.840)
	(99.093,67.683)
	(99.129,70.413)
	(100.000,75.000)

\path(100,75)	(101.405,78.049)
	(103.485,81.321)
	(106.322,83.933)
	(110.000,85.000)

\path(110,85)	(113.677,83.933)
	(116.515,81.321)
	(118.595,78.049)
	(120.000,75.000)

\path(120,75)	(120.581,71.411)
	(120.435,68.675)
	(120.000,65.000)

\put(10,35){\makebox(0,0){$b$}}
\put(10,65){\makebox(0,0){$a$}}
\put(230,55){\makebox(0,0){$a$}}
\put(190,55){\makebox(0,0){$b$}}
\put(80,55){\makebox(0,0){$a$}}
\end{picture}
}
\ee
The other property of the antipode is
\(\Delta\circ Sa = \tau\circ(S\otimes S)\circ\Delta a\). It can be
checked graphically:

\be
\Delta\circ
\raisebox{-0.52in}{
\setlength{\unitlength}{0.0125in}
\begin{picture}(141,85)(0,-10)
\thicklines
\path(31,45)(31,25)(11,25)
	(11,45)(31,45)
\path(41,10)(41,40)
\path(43.000,32.000)(41.000,40.000)(39.000,32.000)
\path(1,40)(1,65)
\path(41,10)(41,5)
\path(116,45)(116,25)(96,25)
	(96,45)(116,45)
\path(86,40)(86,55)(101,70)
\path(56,32)(71,32)
\path(56,38)(71,38)
\path(86,55)(71,70)
\put(106,35){\makebox(0,0){$a$}}
\path(111,0)(126,15)(126,40)
\path(128.000,32.000)(126.000,40.000)(124.000,32.000)
\path(126,15)(141,0)
\path(21,45)	(20.565,48.675)
	(20.419,51.411)
	(21.000,55.000)

\path(21,55)	(22.405,58.049)
	(24.485,61.322)
	(27.322,63.933)
	(31.000,65.000)

\path(31,65)	(34.677,63.933)
	(37.515,61.321)
	(39.595,58.049)
	(41.000,55.000)

\path(41,55)	(41.435,52.707)
	(41.387,49.920)
	(41.000,45.000)

\path(41,45)	(41.000,43.266)
	(41.000,40.000)

\path(21,25)	(21.435,21.325)
	(21.581,18.589)
	(21.000,15.000)

\path(21,15)	(19.595,11.951)
	(17.515,8.679)
	(14.678,6.067)
	(11.000,5.000)

\path(11,5)	(7.322,6.067)
	(4.485,8.679)
	(2.405,11.951)
	(1.000,15.000)

\path(1,15)	(0.565,17.293)
	(0.613,20.080)
	(1.000,25.000)

\path(1,25)	(1.000,27.203)
	(1.000,30.203)
	(1.000,34.352)
	(1.000,36.966)
	(1.000,40.000)

\path(3.000,32.000)(1.000,40.000)(-1.000,32.000)
\path(106,25)	(106.000,21.734)
	(106.000,20.000)

\path(106,20)	(106.194,17.540)
	(106.000,15.000)

\path(106,15)	(104.595,11.951)
	(102.515,8.678)
	(99.678,6.067)
	(96.000,5.000)

\path(96,5)	(92.322,6.067)
	(89.485,8.679)
	(87.405,11.951)
	(86.000,15.000)

\path(86,15)	(85.565,17.293)
	(85.613,20.080)
	(86.000,25.000)

\path(86,25)	(86.000,27.203)
	(86.000,30.203)
	(86.000,34.352)
	(86.000,36.966)
	(86.000,40.000)

\path(88.000,32.000)(86.000,40.000)(84.000,32.000)
\path(106,45)	(106.000,48.266)
	(106.000,50.000)

\path(106,50)	(105.806,52.460)
	(106.000,55.000)

\path(106,55)	(107.405,58.049)
	(109.485,61.321)
	(112.322,63.933)
	(116.000,65.000)

\path(116,65)	(119.677,63.933)
	(122.515,61.321)
	(124.595,58.049)
	(126.000,55.000)

\path(126,55)	(126.435,52.707)
	(126.387,49.920)
	(126.000,45.000)

\path(126,45)	(126.000,43.266)
	(126.000,40.000)

\put(21,35){\makebox(0,0){$a$}}
\end{picture}
}
\mbox{\Huge =}
\raisebox{-0.67in}{
\setlength{\unitlength}{0.0125in}
\begin{picture}(105,107)(0,-10)
\thicklines
\path(50,26)(50,36)
\path(50,56)(50,66)
\path(50,26)	(46.760,21.889)
	(44.170,19.009)
	(40.000,16.000)

\path(40,16)	(37.707,15.565)
	(34.920,15.613)
	(30.000,16.000)

\path(30,16)	(27.577,15.636)
	(25.000,16.000)

\path(25,16)	(22.508,19.122)
	(21.128,23.275)
	(20.434,27.541)
	(20.000,31.000)

\path(20,31)	(19.841,33.220)
	(19.859,36.015)
	(20.000,41.000)

\path(20,41)	(20.000,43.476)
	(20.000,46.508)
	(20.000,49.911)
	(20.000,53.500)
	(20.000,57.089)
	(20.000,60.492)
	(20.000,63.524)
	(20.000,66.000)

\path(20,66)	(20.000,68.938)
	(20.000,72.938)
	(20.000,75.483)
	(20.000,78.469)
	(20.000,81.955)
	(20.000,86.000)

\path(60,36)(60,56)(40,56)
	(40,36)(60,36)
\path(50,66)	(45.009,70.422)
	(41.835,74.161)
	(40.243,77.570)
	(40.000,81.000)

\path(40,81)	(41.274,84.450)
	(43.863,87.313)
	(47.020,89.519)
	(50.000,91.000)

\path(50,91)	(52.293,91.435)
	(55.080,91.387)
	(60.000,91.000)

\path(60,91)	(62.203,91.000)
	(65.000,91.000)
	(67.797,91.000)
	(70.000,91.000)

\path(70,91)	(74.985,91.141)
	(77.780,91.159)
	(80.000,91.000)

\path(80,91)	(83.446,90.471)
	(87.689,89.620)
	(91.838,88.208)
	(95.000,86.000)

\path(95,86)	(96.654,83.508)
	(97.850,80.480)
	(98.669,77.094)
	(99.194,73.528)
	(99.507,69.958)
	(99.691,66.561)
	(99.828,63.516)
	(100.000,61.000)

\path(100,61)	(100.144,57.686)
	(100.128,53.491)
	(100.048,49.302)
	(100.000,46.000)

\path(100,46)	(100.000,43.269)
	(100.000,40.125)
	(100.000,36.449)
	(100.000,32.124)
	(100.000,29.683)
	(100.000,27.035)
	(100.000,24.166)
	(100.000,21.062)
	(100.000,17.708)
	(100.000,14.090)
	(100.000,10.192)
	(100.000,6.000)

\put(50,46){\makebox(0,0){$a$}}
\path(50,66)	(53.240,70.111)
	(55.830,72.991)
	(60.000,76.000)

\path(60,76)	(62.293,76.435)
	(65.080,76.387)
	(70.000,76.000)

\path(70,76)	(72.423,76.364)
	(75.000,76.000)

\path(75,76)	(77.492,72.878)
	(78.872,68.725)
	(79.566,64.459)
	(80.000,61.000)

\path(80,61)	(80.159,58.780)
	(80.141,55.985)
	(80.000,51.000)

\path(80,51)	(80.000,48.524)
	(80.000,45.492)
	(80.000,42.089)
	(80.000,38.500)
	(80.000,34.911)
	(80.000,31.508)
	(80.000,28.476)
	(80.000,26.000)

\path(80,26)	(80.000,23.062)
	(80.000,19.062)
	(80.000,16.517)
	(80.000,13.531)
	(80.000,10.045)
	(80.000,6.000)

\path(50,26)	(54.991,21.578)
	(58.165,17.838)
	(59.757,14.429)
	(60.000,11.000)

\path(60,11)	(58.726,7.550)
	(56.137,4.688)
	(52.980,2.481)
	(50.000,1.000)

\path(50,1)	(47.707,0.565)
	(44.920,0.613)
	(40.000,1.000)

\path(40,1)	(37.797,1.000)
	(35.000,1.000)
	(32.203,1.000)
	(30.000,1.000)

\path(30,1)	(25.015,0.859)
	(22.220,0.841)
	(20.000,1.000)

\path(20,1)	(16.556,1.550)
	(12.316,2.437)
	(8.168,3.855)
	(5.000,6.000)

\path(5,6)	(2.553,10.344)
	(1.752,13.042)
	(1.166,15.897)
	(0.747,18.765)
	(0.445,21.502)
	(0.000,26.000)

\path(0,26)	(-0.239,30.427)
	(-0.249,33.150)
	(-0.213,36.018)
	(-0.150,38.884)
	(-0.080,41.600)
	(0.000,46.000)

\path(0,46)	(0.000,48.731)
	(0.000,51.875)
	(0.000,55.551)
	(0.000,59.876)
	(0.000,62.317)
	(0.000,64.965)
	(0.000,67.834)
	(0.000,70.938)
	(0.000,74.292)
	(0.000,77.910)
	(0.000,81.808)
	(0.000,86.000)

\end{picture}
}
\mbox{\Large =\ } \tau\circ(S\otimes S)
\raisebox{-0.6in}{
\setlength{\unitlength}{0.0125in}
\begin{picture}(30,95)(0,-10)
\thicklines
\path(15,50)(15,65)
\path(15,30)(15,15)
\path(25,30)(25,50)(5,50)
	(5,30)(25,30)
\path(0,0)(15,15)(30,0)
\put(15,40){\makebox(0,0){$a$}}
\path(0,80)(15,65)(30,80)
\end{picture}
}
\ee
where $\tau$ is the flip operator: $V_{j_1}\otimes V_{j_2}
 \stackrel{\tau}{\longrightarrow} V_{j_2}\otimes V_{j_1}$.

Having equipped $\al$ with an $R$-matrix, we obtain a quasi-triangular
Hopf algebra $(\al,R)$. The $R$-matrix obeys the Yang-Baxter equation
(\ref{YB}). In our context, it will be more convenient to consider the
$\r$-matrix

\be
\r := \tau \circ R
\ee
which can be represented graphically as

\be
\bigg\{\r=\tau \circ \sum_i \alpha_i\otimes\beta_i\bigg\}\cong
\raisebox{-1cm}{\bp(20,20)
\thicklines
\put(10,10){\line(1,1){7}}
\put(10,10){\line(-1,-1){7}}
\put(11,9){\line(1,-1){6}}
\put(9,11){\line(-1,1){6}}
\ep}
\ee
It is invertible
\be
\bigg\{\r^{-1}=\tau\circ\sum_i \beta_i\otimes S(\alpha_i)\bigg\}\cong
\Bigg\{
\raisebox{-0.4in}{
\setlength{\unitlength}{0.0125in}

}\Bigg\}
\ee

A Hopf algebra is called {\em triangular} if $\r^2=1$.

The standard definition of the $\r$-matrix requires that the following
properties hold

\be
\bigg\{(\Delta\otimes id)\r = \r_{12}\r_{23}\bigg\}\cong\Bigg\{
(\Delta\otimes id)
\raisebox{-0.4in}{
\setlength{\unitlength}{0.0125in}
%
}
\Bigg\}
\ee
along with the general form of \eq{RTT=TTR}

\be
\bigg\{\r(\Delta a)\r^{-1}=\Delta a,\ \forall a \in\al\bigg\}\cong
\bigg\{
\raisebox{-0.75in}{
\setlength{\unitlength}{0.0125in}
%
}\ \bigg\}
\ee

These equations are equivalent to the Yang-Baxter one

\be
\bigg\{\r_{12}\r_{23}\r_{12}=\r_{23}\r_{12}\r_{23}\bigg\}\cong
\bigg\{
\raisebox{-0.5in}{
\setlength{\unitlength}{0.0125in}
%
}\bigg\}
\ee

The standard relations including the antipode are

\be
\bigg\{(S\otimes id)\r = \r^{-1}\bigg\}\cong \bigg\{
(S\otimes id)
\raisebox{-0.4in}{
\setlength{\unitlength}{0.0125in}
%
}
\bigg\}
\ee
and those with the co-unit are

\be
\bigg\{(\eps\otimes id)\r = (id\otimes\eps)\r=1\bigg\}\cong \bigg\{
\raisebox{-0.4in}{
\setlength{\unitlength}{0.0125in}
%
}
\bigg\}
\ee

The standard way to introduce the ribbon Hopf algebra structure
on $\al$ is to bring forward the element $u\in\al$
defined as \cite{ReshTur}

\be
\bigg\{u:=\sum_i S(\beta_i)\alpha_i\bigg\}\cong 
\raisebox{-0.4in}{
\setlength{\unitlength}{0.0125in}
\thicklines
%
}
\ee
The element $v^2=uS(u)$ lies in the center of $\al$. 

A {\em ribbon Hopf algebra} $\Uq=(\al,\r,v)$
is a quasi-triangular Hopf algebra $(\al,\r)$ equipped with a
central invertible element $v\in\al$,

\be
v:=
\raisebox{-0.4in}{
\setlength{\unitlength}{0.0125in}
%
}
\ee
obeying the properties: $\eps(v)=1$,

\be\ba{l}
\bigg\{uS(u)= v^2\bigg\}\cong \Bigg\{u\circ\Bigg(
\raisebox{-0.3in}{
\setlength{\unitlength}{0.0125in}
%
}
\ee
allows for defining the q-trace of an operator

\be
\qtr(a) := \tr(auv^{-1})\cong\ 
\raisebox{-0.5in}{
\setlength{\unitlength}{0.0125in}
%
}
\ee
In the tensor square of spaces it takes the form

\be
\bigg\{\tr[a \otimes b \circ \Delta(uv^{-1})]=
\qtr(a)\;\qtr(b)\bigg\}\cong
\Bigg\{
\raisebox{-0.8in}{
\setlength{\unitlength}{0.0125in}
%
}
\Bigg\}
\ee
Following Reshetikhin and Turaev, we shall call this operation the
{\em closing} of a tangle.

The quantum dimension of a module, $V_j$, is, by definition, the
q-trace of the identity operator:

\be
d_j := \qtr (1_{V^j})\cong
\raisebox{-0.4in}{
\setlength{\unitlength}{0.0125in}
%
}
\ee

\subsection{Algebra of fields and gauge invariance}

In Section~2.2 we have described a Heegaard diagram as a handlebody with
a given system of $\alpha$-cycles and characteristic curves on its
boundary.  Every $\alpha$-cycle span a disk $D$ in a handlebody $H$. The
disk can be thickened to a plate $P$. In this way we obtain a collection of
disjoint plates in $H$ ($P_i\bigcap P_j = \emptyset$, if $i\neq j$). Each
plate corresponds to a 1-cell of a base cell complex $C$ from which the
Heegaard diagram has been read off. By detaching the plates, $H$
reassembles into a collection of 3-balls $\{B\}$, each corresponding to a
0-cell of $C$.

\noindent
{\sc Definition~1}: We construct qQCD$\mbox{}_3$ functor in the following way:
\begin{enumerate}
\item A gauge variable taking values in a ribbon QUE algebra $\Uq$
is put into correspondence to each plate:
\[P_k \longrightarrow a_k\in \Uq \]
The variables attached to different plates are distinct elements of $\Uq$,
hence their matrix elements are co-commutative. 
\item All the characteristic curves are colored with irreducible
finite-dimensional representations of $\Uq$.
\item If on a boundary of the $k$-th plate there are $n_k$ disjoint cuts of
the characteristic curves colored with representations
$j_1,j_2,\ldots,j_{n_k}$, we construct a {\em gauge field tangle} by
repeatedly applying the co-multiplication:

\[
F_k=\Delta^{n_k-1}(a_k):\ V_{j_1}\otimes\ldots\otimes V_{j_{n_k}}\to
V_{j_1}\otimes\ldots\otimes V_{j_{n_k}}
\]
or, graphically,

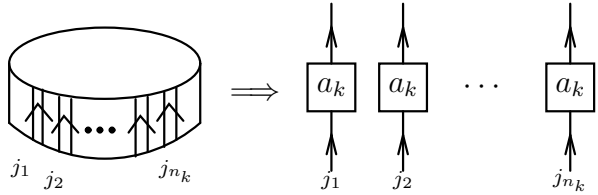
\[
\setlength{\unitlength}{0.0125in}
\begin{picture}(270,89)(0,-10)
\thicklines
\put(33,21){\blacken\ellipse{2}{2}}
\put(33,21){\ellipse{2}{2}}
\put(38,21){\blacken\ellipse{2}{2}}
\put(38,21){\ellipse{2}{2}}
\put(43,21){\blacken\ellipse{2}{2}}
\put(43,21){\ellipse{2}{2}}
\put(40,49){\ellipse{80}{30}}
\path(0,49)(0,24)
\path(80,49)(80,24)
\path(10,39)(10,17)
\path(14,38)(14,15)
\path(21,35)(21,12)
\path(26,34)(26,11)
\path(65,37)(65,14)
\path(58,35)(58,12)
\path(53,34)(53,10)
\path(70,39)(70,17)
\path(145,49)(145,29)(125,29)
	(125,49)(145,49)
\path(175,49)(175,29)(155,29)
	(155,49)(175,49)
\path(245,49)(245,29)(225,29)
	(225,49)(245,49)
\path(135,49)(135,64)
\path(137.000,56.000)(135.000,64.000)(133.000,56.000)
\path(165,49)(165,64)
\path(167.000,56.000)(165.000,64.000)(163.000,56.000)
\path(235,49)(235,64)
\path(237.000,56.000)(235.000,64.000)(233.000,56.000)
\put(40.000,69.833){\arc{121.666}{0.8533}{2.2883}}
\path(135,4)(135,19)
\path(137.000,11.000)(135.000,19.000)(133.000,11.000)
\path(165,4)(165,19)
\path(167.000,11.000)(165.000,19.000)(163.000,11.000)
\path(235,4)(235,19)
\path(237.000,11.000)(235.000,19.000)(233.000,11.000)
\path(135,19)(135,29)
\path(165,19)(165,29)
\path(235,19)(235,29)
\path(135,64)(135,74)
\path(165,64)(165,74)
\path(235,64)(235,74)
\path(18,20)(23,27)(29,20)
\path(62,25)(67,32)(73,25)
\path(7,25)(12,32)(18,25)
\path(50,20)(55,27)(61,20)
\put(19,0){\makebox(0,0){$\scr j_2$}}
\put(5,5){\makebox(0,0){$\scr \scr j_1$}}
\put(70,4){\makebox(0,0){$\scr j_{n_k}$}}
\put(103,36){\makebox(0,0){$\Longrightarrow$}}
\put(190,39){\makebox(0,0)[l]{$\cdots$}}
\put(135,39){\makebox(0,0){$a_k$}}
\put(165,39){\makebox(0,0){$a_k$}}
\put(235,39){\makebox(0,0){$a_k$}}
\put(135,0){\makebox(0,0){$\scr j_1$}}
\put(165,0){\makebox(0,0){$\scr j_2$}}
\put(235,0){\makebox(0,0){$\scr j_{n_k}$}}
\end{picture}
\]

One has to respect the cyclic order and mutual orientations of the cuts. A
reversion of an orientation of a cut corresponds to the 
conjugation\mbox{}$^($\footnote{\mbox{}$^)$ with respect to some fixed
$\ast$-structure.}\mbox{}$^)$ of the corresponding matrix element.
\item One puts into correspondence to each ball $B_i\in\{B\}$ carrying a
pattern of the characteristic curves on its boundary a {\em vertex tangle}
by using the Reshetikhin-Turaev functor {\bf ctang \(\to\) rep\(_{\Uq}\)}.
\item In the end, the pieces are attached together. For it, one embeds the
handlebody into $R^3$ in such a way that the cuts of the characteristic
curves on boundaries of the plates project to distinct points on the
$(x,y)$ plane and onto disjoint segments on the $(x,z)$ plane. Then one can
use the $(x,z)$ projection of the vertex tangles to complete the
construction in terms of elements of a ribbon Hopf algebra as was
depicted in the previous section. A result is a functional taking values
in $C$.
$\Box$
\end{enumerate}

\noindent
{\em Remarks}:~1)~Let us notice that the initial data are colorings and
directions of the characteristic curves as well as $\Uq$ elements attached
to the plates.\\ 
2)~One can sum over all the colorings with arbitrary
weights as in \eq{Z}. As we are restricted to real forms of $\Uq(sl(n))$,
the result has to be independent of the directions of the characteristic
curves.\\ 
3)~Modules appearing in different vertex tangles are independent.
However, after the assemblage, all pieces are fit together and, permuting
matrix elements adjacent via a vertex tangle, one has to deform the tangle,
which means some effective non-cocommutativity. We shall dwell at this
point later. Now, let us simply illustrate what may happen by the example
shown in Figure~1.\\ 
4)~Clearly, the construction gives different results
for non-isotopic embeddings of $H$ into $R^3$. This lack of
self-consistency will disappear after the integration over gauge fields
(see the next section).

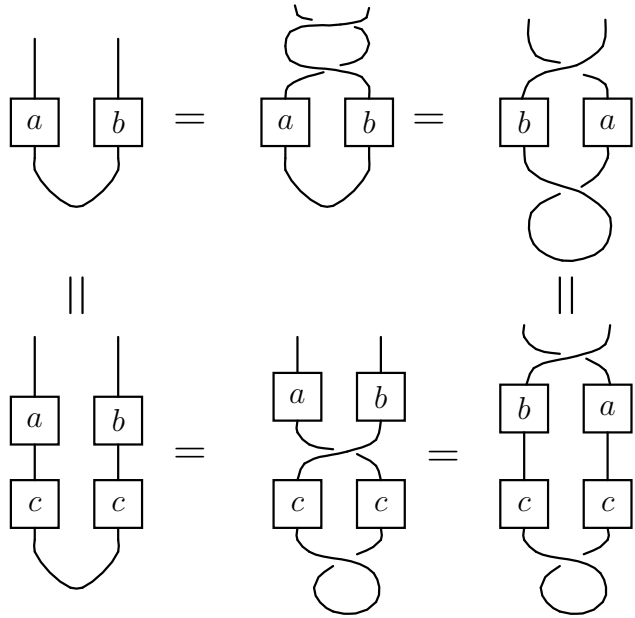
\begin{figure}
\begin{center}
\setlength{\unitlength}{0.0125in}
\begin{picture}(268,270)(0,-10)
\thicklines
\path(55,216)(55,196)(35,196)
	(35,216)(55,216)
\path(10,216)(10,241)
\path(45,216)(45,241)
\path(125,216)(125,196)(105,196)
	(105,216)(125,216)
\path(160,216)(160,196)(140,196)
	(140,216)(160,216)
\path(225,216)(225,196)(205,196)
	(205,216)(225,216)
\path(260,216)(260,196)(240,196)
	(240,216)(260,216)
\path(25,141)(25,126)
\path(30,141)(30,126)
\path(230,141)(230,126)
\path(235,141)(235,126)
\path(10,91)(10,116)
\path(45,91)(45,116)
\path(55,91)(55,71)(35,71)
	(35,91)(55,91)
\path(20,91)(20,71)(0,71)
	(0,91)(20,91)
\path(55,56)(55,36)(35,36)
	(35,56)(55,56)
\path(20,56)(20,36)(0,36)
	(0,56)(20,56)
\path(10,56)(10,71)
\path(45,56)(45,71)
\path(130,101)(130,81)(110,81)
	(110,101)(130,101)
\path(130,56)(130,36)(110,36)
	(110,56)(130,56)
\path(165,56)(165,36)(145,36)
	(145,56)(165,56)
\path(165,101)(165,81)(145,81)
	(145,101)(165,101)
\path(120,101)(120,116)
\path(155,101)(155,116)
\path(225,96)(225,76)(205,76)
	(205,96)(225,96)
\path(225,56)(225,36)(205,36)
	(205,56)(225,56)
\path(260,56)(260,36)(240,36)
	(240,56)(260,56)
\path(260,96)(260,76)(240,76)
	(240,96)(260,96)
\path(215,76)(215,56)
\path(250,76)(250,56)
\path(10,196)	(10.000,192.734)
	(10.000,191.000)

\path(10,191)	(9.806,188.540)
	(10.000,186.000)

\path(10,186)	(12.524,181.734)
	(16.339,177.339)
	(20.734,173.524)
	(25.000,171.000)

\path(25,171)	(27.500,170.613)
	(30.000,171.000)

\path(30,171)	(34.266,173.524)
	(38.661,177.339)
	(42.476,181.734)
	(45.000,186.000)

\path(45,186)	(45.194,188.540)
	(45.000,191.000)

\path(45,191)	(45.000,192.734)
	(45.000,196.000)

\path(115,216)	(115.328,219.262)
	(116.000,221.000)

\path(116,221)	(118.382,222.786)
	(121.000,224.000)

\path(121,224)	(124.434,225.207)
	(127.208,225.995)
	(131.000,227.000)

\path(126,249)	(122.027,249.791)
	(120.000,251.000)

\path(120,251)	(119.000,255.000)

\path(138,230)	(141.438,230.711)
	(143.944,231.401)
	(147.000,233.000)

\path(147,233)	(149.013,235.725)
	(150.000,239.000)

\path(150,239)	(149.013,242.275)
	(147.000,245.000)

\path(147,245)	(144.000,246.000)

\path(150,216)	(150.228,219.223)
	(150.000,221.000)

\path(150,221)	(147.000,225.000)

\path(147,225)	(144.612,226.079)
	(141.537,226.890)
	(138.444,227.506)
	(136.000,228.000)

\path(136,228)	(131.828,228.517)
	(129.246,228.698)
	(126.520,228.932)
	(123.792,229.301)
	(121.202,229.882)
	(117.000,232.000)

\path(117,232)	(115.468,234.868)
	(115.000,238.000)

\path(115,238)	(115.436,241.136)
	(117.000,244.000)

\path(117,244)	(121.742,246.108)
	(124.609,246.541)
	(127.610,246.711)
	(130.597,246.729)
	(133.423,246.709)
	(138.000,247.000)

\path(138,247)	(140.457,247.345)
	(143.551,247.772)
	(146.620,248.564)
	(149.000,250.000)

\path(149,250)	(150.000,254.000)

\path(115,196)	(115.000,192.734)
	(115.000,191.000)

\path(115,191)	(114.806,188.540)
	(115.000,186.000)

\path(115,186)	(117.524,181.734)
	(121.339,177.339)
	(125.734,173.524)
	(130.000,171.000)

\path(130,171)	(132.500,170.613)
	(135.000,171.000)

\path(135,171)	(139.266,173.524)
	(143.661,177.339)
	(147.476,181.734)
	(150.000,186.000)

\path(150,186)	(150.193,188.540)
	(150.000,191.000)

\path(150,191)	(150.000,192.734)
	(150.000,196.000)

\path(20,216)(20,196)(0,196)
	(0,216)(20,216)
\path(215,196)	(215.161,192.245)
	(215.516,189.483)
	(217.000,186.000)

\path(217,186)	(221.827,182.497)
	(224.981,181.190)
	(228.358,180.076)
	(231.752,179.077)
	(234.960,178.116)
	(237.777,177.117)
	(240.000,176.000)

\path(240,176)	(242.768,174.242)
	(246.096,171.901)
	(249.126,169.111)
	(251.000,166.000)

\path(251,166)	(251.431,162.754)
	(250.993,159.155)
	(249.808,155.729)
	(248.000,153.000)

\path(248,153)	(244.465,150.488)
	(239.918,148.816)
	(235.163,147.986)
	(231.000,148.000)

\path(231,148)	(227.840,148.805)
	(224.394,150.388)
	(221.251,152.527)
	(219.000,155.000)

\path(219,155)	(217.751,157.886)
	(217.096,161.371)
	(217.143,164.921)
	(218.000,168.000)

\path(218,168)	(221.701,171.962)
	(225.135,173.878)
	(230.000,176.000)

\path(239,179)	(243.688,182.043)
	(246.000,184.000)

\path(246,184)	(248.296,187.320)
	(250.000,191.000)

\path(250,191)	(250.178,192.763)
	(250.000,196.000)

\path(249,249)	(249.164,244.923)
	(249.068,241.903)
	(248.000,238.000)

\path(248,238)	(245.913,235.558)
	(242.911,233.277)
	(239.704,231.358)
	(237.000,230.000)

\path(237,230)	(234.154,229.133)
	(230.501,228.430)
	(226.848,227.762)
	(224.000,227.000)

\path(224,227)	(220.358,225.321)
	(217.000,223.000)

\path(217,223)	(215.613,220.649)
	(214.000,216.000)

\path(230,231)	(226.193,231.644)
	(223.414,232.312)
	(220.000,234.000)

\path(220,234)	(218.115,236.848)
	(217.000,240.000)

\path(217,240)	(216.718,243.163)
	(216.789,245.642)
	(217.000,249.000)

\path(250,216)	(249.834,219.893)
	(249.000,222.000)

\path(249,222)	(246.036,224.160)
	(243.507,225.073)
	(240.000,226.000)

\path(10,36)	(10.000,32.734)
	(10.000,31.000)

\path(10,31)	(9.806,28.540)
	(10.000,26.000)

\path(10,26)	(12.524,21.734)
	(16.339,17.339)
	(20.734,13.524)
	(25.000,11.000)

\path(25,11)	(27.500,10.613)
	(30.000,11.000)

\path(30,11)	(34.266,13.524)
	(38.661,17.339)
	(42.476,21.734)
	(45.000,26.000)

\path(45,26)	(45.194,28.540)
	(45.000,31.000)

\path(45,31)	(45.000,32.734)
	(45.000,36.000)

\path(120,36)	(119.710,32.794)
	(120.000,31.000)

\path(120,31)	(122.227,28.178)
	(125.000,26.000)

\path(125,26)	(127.616,24.864)
	(130.908,24.065)
	(134.646,23.438)
	(138.601,22.813)
	(142.543,22.024)
	(146.243,20.904)
	(149.472,19.285)
	(152.000,17.000)

\path(152,17)	(153.521,14.348)
	(154.362,11.190)
	(154.272,7.938)
	(153.000,5.000)

\path(153,5)	(149.272,1.830)
	(144.539,0.261)
	(142.025,0.005)
	(139.536,0.062)
	(135.000,1.000)

\path(135,1)	(132.148,2.477)
	(129.546,4.780)
	(127.671,7.693)
	(127.000,11.000)

\path(127,11)	(128.813,15.389)
	(131.266,17.561)
	(135.000,20.000)

\path(155,36)	(155.415,32.826)
	(155.000,31.000)

\path(155,31)	(151.846,27.980)
	(149.001,26.555)
	(145.000,25.000)

\path(120,56)	(121.435,60.674)
	(123.000,63.000)

\path(123,63)	(126.917,65.047)
	(129.373,65.737)
	(131.972,66.276)
	(134.575,66.718)
	(137.040,67.117)
	(141.000,68.000)

\path(141,68)	(143.720,68.794)
	(147.144,69.806)
	(150.496,71.165)
	(153.000,73.000)

\path(153,73)	(154.304,75.756)
	(154.704,77.974)
	(155.000,81.000)

\path(120,81)	(119.710,77.794)
	(120.000,76.000)

\path(120,76)	(122.186,73.154)
	(125.000,71.000)

\path(125,71)	(128.459,69.882)
	(131.227,69.435)
	(135.000,69.000)

\path(155,56)	(154.011,59.960)
	(153.000,62.000)

\path(153,62)	(150.396,64.179)
	(148.141,65.451)
	(145.000,67.000)

\path(215,36)	(214.710,32.794)
	(215.000,31.000)

\path(215,31)	(217.227,28.178)
	(220.000,26.000)

\path(220,26)	(222.616,24.864)
	(225.908,24.065)
	(229.646,23.438)
	(233.601,22.813)
	(237.543,22.024)
	(241.243,20.904)
	(244.472,19.285)
	(247.000,17.000)

\path(247,17)	(248.521,14.348)
	(249.362,11.190)
	(249.272,7.938)
	(248.000,5.000)

\path(248,5)	(244.272,1.830)
	(239.539,0.261)
	(237.025,0.005)
	(234.536,0.062)
	(230.000,1.000)

\path(230,1)	(227.148,2.477)
	(224.546,4.780)
	(222.671,7.693)
	(222.000,11.000)

\path(222,11)	(223.813,15.389)
	(226.266,17.561)
	(230.000,20.000)

\path(250,36)	(250.415,32.826)
	(250.000,31.000)

\path(250,31)	(246.846,27.980)
	(244.001,26.555)
	(240.000,25.000)

\path(216,96)	(217.435,100.674)
	(219.000,103.000)

\path(219,103)	(222.917,105.047)
	(225.373,105.737)
	(227.972,106.276)
	(230.575,106.718)
	(233.040,107.117)
	(237.000,108.000)

\path(237,108)	(239.720,108.794)
	(243.144,109.806)
	(246.496,111.165)
	(249.000,113.000)

\path(249,113)	(250.304,115.756)
	(250.704,117.974)
	(251.000,121.000)

\path(251,96)	(250.011,99.960)
	(249.000,102.000)

\path(249,102)	(246.396,104.179)
	(244.141,105.451)
	(241.000,107.000)

\path(215,121)	(214.710,117.794)
	(215.000,116.000)

\path(215,116)	(217.186,113.154)
	(220.000,111.000)

\path(220,111)	(223.459,109.882)
	(226.227,109.435)
	(230.000,109.000)

\put(75,206){\makebox(0,0){\Large =}}
\put(175,206){\makebox(0,0){\Large =}}
\put(215,206){\makebox(0,0){$b$}}
\put(250,206){\makebox(0,0){$a$}}
\put(115,206){\makebox(0,0){$a$}}
\put(150,206){\makebox(0,0){$b$}}
\put(45,206){\makebox(0,0){$b$}}
\put(10,206){\makebox(0,0){$a$}}
\put(10,81){\makebox(0,0){$a$}}
\put(45,81){\makebox(0,0){$b$}}
\put(10,46){\makebox(0,0){$c$}}
\put(45,46){\makebox(0,0){$c$}}
\put(75,66){\makebox(0,0){\Large =}}
\put(120,91){\makebox(0,0){$a$}}
\put(120,46){\makebox(0,0){$c$}}
\put(155,46){\makebox(0,0){$c$}}
\put(155,91){\makebox(0,0){$b$}}
\put(181,65){\makebox(0,0){\Large =}}
\put(215,46){\makebox(0,0){$c$}}
\put(250,46){\makebox(0,0){$c$}}
\put(215,86){\makebox(0,0){$b$}}
\put(250,86){\makebox(0,0){$a$}}
\end{picture}
\end{center}
\caption{An illustration of the non-cocommutativity of fields adjacent via a
vertex.}
\end{figure}

Let us now discuss the issue of gauge invariance within our framework. In
the general settings of gauge theory the basic object is a fiber bundle
over a base manifold. Gauge field performs a parallel transport of fibers
and thus is interpreted as a $G$-connection in sections of the bundle. Gauge
transformations act by automorphisms of the fibers. To make it explicit,
one has to choose some $G$-basis at each point of the base. A quantity is
gauge invariant if it is independent of a particular choice of the
bases.

In LGT a base manifold is substituted by a finite cell
complex. Therefore, instead of a fiber bundle, one has a tensor product
of $G$-modules, one for each 0-cell in the complex.  In our construction of
qQCD$_3$, 0-cells are associated with the vertex tangles. One chooses bases
of the $\Uq$-modules for each tangle independently and then sandwiches the
matrix elements between them. Thus, we can reformulate gauge invariance as
the requirement of independence from particular choices of all the
bases. As in the quantum case the notion of the group manifold is absent,
one cannot translate a change of a frame into a group rotation.  
Although these changes can be given a matrix form, their status is quite
different from the one of gauge fields. Therefore, 
we lose a contact with the explicit formula (\ref{gtrans}).

\subsection{Partition function}

We connect integration with the co-unit. By definition, it is a linear
functional projecting onto the trivial representation. For the matrix
elements, we have

\be
\bigg\{\int da\; D^j(a) = \delta_{j,0}\bigg\}\cong\bigg\{
\eps(\raisebox{-0.31in}{
\setlength{\unitlength}{0.0125in}
\begin{picture}(6,55)(5,-9)
\thicklines
\path(5,25)(5,40)
\path(10,25)(10,15)(0,15)
	(0,25)(10,25)
\path(5,15)(5,0)
\put(8,-2){\makebox(0,0){$\scr j$}}
\put(5,20){\makebox(0,0){$a$}}
\end{picture}
}\;)=\delta_{j,0}\bigg\}
\label{int}
\ee
An integral of an arbitrary product of matrix elements can be reduced
to the basic one (\ref{int}) by subsequently applying the tensor
product decomposition with the Clebsch-Gordan coefficients. For
example, the orthogonality of matrix elements reads

\be
\eps(
\setlength{\unitlength}{0.0125in}
\raisebox{-.48in}{
\begin{picture}(35,79)(0,-10)
\thicklines
\path(5,35)(5,50)
\path(5,25)(5,10)
\path(25,35)(25,25)(15,25)
	(15,35)(25,35)
\path(20,25)(20,10)
\path(10,35)(10,25)(0,25)
	(0,35)(10,35)
\path(20,35)(20,50)
\put(2,2){\makebox(0,0){$\scr j_1\alpha_1$}}
\put(23,2){\makebox(0,0){$\scr j_2\alpha_2$}}
\put(5,30){\makebox(0,0){$a$}}
\put(20,30){\makebox(0,0){$a$}}
\put(5,55){\makebox(0,0){$\scr\beta_1$}}
\put(20,55){\makebox(0,0){$\scr\beta_2$}}
\end{picture}}
\!\!) = \sum \eps(
\raisebox{-0.48in}{
\setlength{\unitlength}{0.0125in}
\begin{picture}(30,79)(0,-10)
\thicklines
\path(10,35)(10,45)
\path(10,25)(10,15)
\path(0,55)(10,45)(20,55)
\path(15,35)(15,25)(5,25)
	(5,35)(15,35)
\path(0,5)(10,15)(20,5)
\put(10,30){\makebox(0,0){$a$}}
\put(0,60){\makebox(0,0){$\scr\beta_1$}}
\put(17,61){\makebox(0,0){$\scr\beta_2$}}
\put(0,0){\makebox(0,0){$\scr j_1\alpha_1$}}
\put(20,0){\makebox(0,0){$\scr j_2\alpha_2$}}
\end{picture}
}) =
\frac{\delta_{j_1,j_2}}{d_{j_1}}
\raisebox{-0.48in}{
\setlength{\unitlength}{0.0125in}
\begin{picture}(35,84)(0,-10)
\thicklines
\path(1,10)	(0.565,13.675)
	(0.420,16.411)
	(1.000,20.000)

\path(1,20)	(2.683,22.844)
	(5.226,25.774)
	(8.156,28.318)
	(11.000,30.000)

\path(11,30)	(13.500,30.387)
	(16.000,30.000)

\path(16,30)	(18.843,28.317)
	(21.774,25.774)
	(24.317,22.844)
	(26.000,20.000)

\path(26,20)	(26.581,16.411)
	(26.435,13.675)
	(26.000,10.000)

\path(1,60)	(0.565,56.325)
	(0.420,53.589)
	(1.000,50.000)

\path(1,50)	(2.683,47.156)
	(5.226,44.226)
	(8.157,41.683)
	(11.000,40.000)

\path(11,40)	(13.500,39.613)
	(16.000,40.000)

\path(16,40)	(18.844,41.683)
	(21.774,44.226)
	(24.317,47.156)
	(26.000,50.000)

\path(26,50)	(26.581,53.589)
	(26.435,56.325)
	(26.000,60.000)

\put(24,63){\makebox(0,0){$\scr\beta_2$}}
\put(1,63){\makebox(0,0){$\scr\beta_1$}}
\put(27,0){\makebox(0,0){$\scr j_2\alpha_2$}}
\put(0,0){\makebox(0,0){$\scr j_1\alpha_1$}}
\end{picture}
}
\label{orth}
\ee
where $d_j$ is the q-dimension of a module $V_j$.

The selfconsistency of the definition can be easily checked 

\be
\setlength{\unitlength}{0.0125in}
\raisebox{-0.6in}{\begin{picture}(215,85)(0,-10)
\thicklines
\path(30,0)(30,15)
\path(35,25)(35,15)(25,15)
	(25,25)(35,25)
\path(30,40)(30,25)
\path(35,50)(35,40)(25,40)
	(25,50)(35,50)
\path(30,50)(30,65)
\path(55,25)(55,15)(45,15)
	(45,25)(55,25)
\path(50,0)(50,15)
\path(55,50)(55,40)(45,40)
	(45,50)(55,50)
\path(50,40)(50,25)
\path(50,50)(50,65)
\spline(125,70)
(125,65)(125,60)
	(135,55)(145,60)(145,65)(145,70)
\put(135,35){\ellipse{20}{20}}
\spline(125,0)
(125,5)(125,10)
	(135,15)(145,10)(145,5)(145,0)
\spline(195,10)
(195,15)(195,20)
	(205,25)(215,20)(215,15)(215,10)
\spline(195,65)
(195,60)(195,55)
	(205,50)(215,55)(215,60)(215,65)
\put(160,35){\makebox(0,0)[l]{$\dsty = \frac1{d_j}$}}
\put(30,45){\makebox(0,0){$a$}}
\put(10,45){\makebox(0,0)[l]{$\eps($}}
\put(10,20){\makebox(0,0)[l]{$\eps($}}
\put(30,20){\makebox(0,0){$b$}}
\put(50,20){\makebox(0,0){$b$}}
\put(50,45){\makebox(0,0){$a$}}
\put(60,20){\makebox(0,0){$)$}}
\put(60,45){\makebox(0,0){$)$}}
\put(75,35){\makebox(0,0)[l]{$\dsty =\frac1{d^2_j}$}}
\end{picture}
}
\ee

The main property of the Haar integral is the right/left invariance:

\be
\raisebox{-0.8in}{
\setlength{\unitlength}{0.0125in}
\begin{picture}(320,115)(0,-10)
\thicklines
\path(90,40)(90,60)(60,60)
	(60,40)(90,40)
\path(40,40)(40,20)
\path(40,80)(40,60)
\path(75,80)(75,60)
\path(75,40)(75,20)
\path(180,55)(180,45)(170,45)
	(170,55)(180,55)
\path(175,55)(175,65)
\path(175,45)(175,35)
\path(170,90)(170,80)(160,80)
	(160,90)(170,90)
\path(190,90)(190,80)(180,80)
	(180,90)(190,90)
\path(170,20)(170,10)(160,10)
	(160,20)(170,20)
\path(190,20)(190,10)(180,10)
	(180,20)(190,20)
\path(165,90)(165,100)
\path(185,90)(185,100)
\path(165,10)(165,0)
\path(185,10)(185,0)
\path(240,20)(240,10)(230,10)
	(230,20)(240,20)
\path(260,20)(260,10)(250,10)
	(250,20)(260,20)
\path(260,90)(260,80)(250,80)
	(250,90)(260,90)
\path(240,90)(240,80)(230,80)
	(230,90)(240,90)
\path(235,10)(235,0)
\path(255,10)(255,0)
\path(55,40)(55,60)(25,60)
	(25,40)(55,40)
\path(255,100)(255,90)
\path(235,100)(235,90)
\path(165,20)	(164.710,23.206)
	(165.000,25.000)

\path(165,25)	(166.405,28.049)
	(168.485,31.322)
	(171.322,33.933)
	(175.000,35.000)

\path(175,35)	(178.678,33.933)
	(181.515,31.321)
	(183.595,28.049)
	(185.000,25.000)

\path(185,25)	(185.290,23.206)
	(185.000,20.000)

\path(165,80)	(164.710,76.794)
	(165.000,75.000)

\path(165,75)	(166.405,71.951)
	(168.485,68.679)
	(171.322,66.067)
	(175.000,65.000)

\path(175,65)	(178.678,66.067)
	(181.515,68.679)
	(183.595,71.951)
	(185.000,75.000)

\path(185,75)	(185.290,76.794)
	(185.000,80.000)

\spline(235,80)
(235,75)(235,70)
	(245,65)(255,70)(255,75)(255,80)
\spline(235,20)
(235,25)(235,30)
	(245,35)(255,30)(255,25)(255,20)
\spline(300,75)
(300,70)(300,65)
	(310,60)(320,65)(320,70)(320,75)
\spline(300,25)
(300,30)(300,35)
	(310,40)(320,35)(320,30)(320,25)
\put(265,50){\makebox(0,0)[l]{$\dsty =\frac1{d_j}$}}
\put(10,50){\makebox(0,0)[l]{$\eps($}}
\put(40,50){\makebox(0,0){$abc$}}
\put(75,50){\makebox(0,0){$abc$}}
\put(95,50){\makebox(0,0)[l]{$\dsty )\;=\;\sum\;\eps($}}
\put(175,50){\makebox(0,0){$b$}}
\put(165,85){\makebox(0,0){$a$}}
\put(185,85){\makebox(0,0){$a$}}
\put(165,15){\makebox(0,0){$c$}}
\put(185,15){\makebox(0,0){$c$}}
\put(190,50){\makebox(0,0)[l]{$\dsty )\;=\frac1{d_j}$}}
\put(235,15){\makebox(0,0){$c$}}
\put(255,15){\makebox(0,0){$c$}}
\put(235,85){\makebox(0,0){$a$}}
\put(255,85){\makebox(0,0){$a$}}
\end{picture}
}
\ee
In the general case, the invariance easily follows from the properties
of the Clebsch-Gordan coefficients and the antipode.

The integral should be used with some caution. For example, a reader has
to be aware that 
\be 
\raisebox{-0.3in}{
\setlength{\unitlength}{0.0125in}
\begin{picture}(115,55)(0,-10)
\thicklines
\path(30,25)(30,40)
\path(30,15)(30,0)
\path(50,40)(50,0)
\path(100,25)(100,15)(90,15)
	(90,25)(100,25)
\path(95,15)(95,0)
\path(95,25)(95,40)
\path(35,25)(35,15)(25,15)
	(25,25)(35,25)
\path(105,40)(105,0)
\put(110,20){\makebox(0,0){$\;)$}}
\put(10,20){\makebox(0,0)[l]{$\eps($}}
\put(30,20){\makebox(0,0){$a$}}
\put(40,20){\makebox(0,0){$)$}}
\put(60,20){\makebox(0,0)[l]{$\not\cong\eps($}}
\put(95,20){\makebox(0,0){$a$}}
\end{picture}
}
\ee 
and the r.h.s. of this formula makes no sense. Otherwise, one could
easily arrive at contradictions. Expressions like $\int dx\:dy\:f(x,y)$ are
inadequate in the quantum case. To exclude ambiguity, we shall always
connect the integration with the linear operator acting on a tensor product
of modules constructed with help of the co-multiplication:

\noindent
{\sc definition~2:}
\[
\int\ := \eps\Big(\Delta^n(\ )\Big):\ (V_1\otimes\ldots\otimes V_{n+1}
\stackrel{\Delta^n}{\longrightarrow} V_1\otimes\ldots\otimes V_{n+1})
\stackrel{\eps}{\longrightarrow} C
\]
It can be calculated recursively:

\be
\raisebox{-0.4in}{
\setlength{\unitlength}{0.0125in}
\begin{picture}(190,65)(0,-10)
\thicklines
\path(60,30)(60,20)(50,20)
	(50,30)(60,30)
\path(55,30)(55,45)
\path(75,30)(75,20)(65,20)
	(65,30)(75,30)
\path(70,20)(70,5)
\path(70,30)(70,45)
\path(25,20)(25,5)
\path(25,30)(25,45)
\path(30,30)(30,20)(20,20)
	(20,30)(30,30)
\path(150,30)(150,20)(140,20)
	(140,30)(150,30)
\path(180,30)(180,20)(170,20)
	(170,30)(180,30)
\path(145,20)(145,5)
\path(145,45)(145,30)
\path(175,20)(175,10)
\path(55,20)(55,5)
\path(175,30)(175,40)
\path(185,50)(175,40)(165,50)
\path(165,0)(175,10)(185,0)
\put(175,25){\makebox(0,0){$a$}}
\put(5,25){\makebox(0,0)[l]{$\eps($}}
\put(55,25){\makebox(0,0){$a$}}
\put(70,25){\makebox(0,0){$a$}}
\put(80,25){\makebox(0,0)[l]{$\dsty )=\sum\eps($}}
\put(25,25){\makebox(0,0){$a$}}
\put(34,25){\makebox(0,0)[l]{$\cdots$}}
\put(154,25){\makebox(0,0)[l]{$\cdots$}}
\put(185,25){\makebox(0,0){$\;)$}}
\put(145,25){\makebox(0,0){$a$}}
\end{picture}
}
\label{recurs}
\ee

To complete our construction we need to specify a real form of $\Uq(sl(n))$
with respect to some fixed $\ast$-structure. We are interested in 2 cases:
(i) $\Uq(su(n))$ which makes sense for real $q$ and (ii) $\Uq(sl(n,R))$ for
$|q|=1$. The Hopf $\ast$-algebra $\Uq(su(n))$ has been well
investigated starting from the pioneering works of Woronowicz and Vaksman
and Soibelman \cite{Woron}.  We need the following facts:
\begin{enumerate}
\item There is the one-to-one correspondence between finite dimensional
irreducible representations of $\Uq(su(n))$ and the classical algebra
${\cal U}(su(n))$.
\item The representation ring of $\Uq(su(n))$ spanned by matrix elements of
finite-dimensional irreducible representations can be regarded as the
$q$-deformation of the algebra of regular functions on $SU(N)$ (the quantum
Peter-Weyl theorem). In particular, there exists a $q$-analog of the group
$\delta$-function.
\item There exists a q-analog of the Haar measure.  The matrix elements are
orthogonal with respect to it.  An explicit representation of them can be
given in terms of $q$-special functions. In this case the group integration
is performed with help of the so-called Jackson integral from the
$q$-special function theory.
\end{enumerate}

Now, we are in a position to define the qQCD$_3$ partition function.
We shall denote the number of $k$-cells in a complex as $N_k$.

\noindent
{\sc Definition~3:} We take the construction of the qQCD$_3$ functor 
introduced in the previous section. Then
\begin{enumerate}
\item We color the characteristic curves with $\Uq(su(n))$ irreps:
$\gamma_i\to j_i$, $i=1,\ldots,N_2$.
\item We put into correspondence to every plate $P_k$ ($k=1,\ldots,N_1$)
the integral

\[
P_k\longrightarrow\eps(\Delta^{n_k-1}(a_k)),\ \ a_k\in\Uq(su(n))
\]
\item By applying Eqs.~(\ref{int}), (\ref{orth}) and (\ref{recurs}) we
obtain a collection of closed 3-valent ribbon graphs $\{\tau\}$, the number
of which equals the number of 0-cells in a base cell complex.  By using the
Reshetikhin-Turaev functor, we calculate the quantum
invariant\mbox{}$^($\footnote{$^)$ Often called the generalized Jones
polynomial.}\mbox{}$^)$, $J(\tau_k)$, for each connected component,
$\tau_k$.  Let us denote their product as

\be
Z_{j_1\ldots j_{N_2}}=\prod_{k=1}^{N_0} J(\tau_k)
\label{Zjj}
\ee
\item The partition function equals the sum over all colorings of  the
characteristic curves 

\be
{\cal Z}_{\beta}=\sum_{\{j_1\ldots j_{N_2}\}}\prod_{k=1}^{N_2}\Big(d_{j_k}
e^{-\beta C_{j_k}}\Big) Z_{j_1\ldots j_{N_2}}
\label{parfun}
\ee
where $d_j$ is the quantum dimension and $C_j$ is a second Casimir eigenvalue.
\end{enumerate}

\noindent
{\em Remarks}:~1)~If $q=1$, this definition reduces to the one given in
Eqs.~(\ref{weight}) and (\ref{Z}).\\ 
2)~${\cal Z}_\beta$ is a
gauge invariant quantity in the sense described in the previous
section. Indeed, as any vertex tangle after the integration gives a closed
ribbon 3-valent graph, the choice of a basis attached to it is
irrelevant.\\ 
3)~After the integration, all non-isotopic embeddings of a handlebody $H$ into
$R^3$ become equivalent and the consideration can be restricted to
isotopies of the handlebody itself.\\
4)~If one considers a cell complex dual to a simplicial one,
the ribbon graph invariants $J(\tau_k)$ in \eq{Zjj} coincide with the quantum
6-$j$ symbols in the Racah-Wiegner normalization.

\subsection{The root of unity case}

For applications most interesting is the case when $q$ equals a primitive
root of unity: \mbox{$q^{\ell}=1$}. Then $\Uq(sl(n))$ possesses the real form
$\Uq(sl(n,R))$. This case is rather complicated
technically. One has to work with the restricted specialization
$\Uq^{\rm res}(sl(n))$ of $\Uq(sl(n))$ and the issue of the duality between
the QF and QUE algebras becomes quite subtle. Fortunately, one can go on
with the notion of the modular Hopf algebra \cite{ReshTur}.

\noindent
{\sc definition~4:} Consider a ribbon Hopf algebra $(\al,\r,v)$ equipped with a
distinguished family $\{V_j\}_{j\in{\cal S}}$ of irreducible $\al$-modules
indexed by a finite set $\cal S$ including the trivial representation
$V_0$. $(\al,\r,v)$ is called a {\em modular Hopf algebra} if  the following
requirements are fulfilled:
\begin{enumerate}
\item qdim$\; V_j\neq 0$, $\forall j\in{\cal S}$.
\item  The set $\{V_j\}_{j\in{\cal S}}$ is equipped with an involution
$j\to j^{\ast}$ such that $V_{j^{\ast}}=V^{\ast}_j$ and
$V^{\ast\ast}_j=V_j$. 
\item For any sequence $j_1,\ldots,j_n\in {\cal S}$
\[
V_{j_1}\otimes V_{j_2}\otimes\ldots V_{j_n}= \bigoplus_{i\in {\cal S}} 
V_i^{\oplus m_i}\oplus I,\ \ \ \ m_i\in N
\]
as $\al$-modules and for all $\al$-module endomorphisms, $f$,
of the ideal $I$
\[
\qtr(f)=0
\]
\item Let $s_{ij}$ be the quantum invariant of the Hopf link, two components of
which are colored with irreps $i$ and $j\in {\cal S}$ 
\[
s_{ij}=\qtr\Big[
\raisebox{-0.9cm}{\bp(15,20)
\thicklines
\put(5,0){$\scr j$}
\put(5.5,3){\line(0,1){4}}
\put(5.5,9){\line(0,1){8}}
\put(3,8){\line(1,0){5}}
\put(3,12){\line(1,0){2}}
\put(6,12){\line(1,0){2}}
\put(2,10){\oval(4,4)[l]}
\put(6.5,10){\oval(4,4)[r]}
\put(11,11){$\scr i$}
\ep}\Big]
\]
then the matrix
$(s_{ij})_{i,j\in{\cal S}}$ is invertible. $\Box$
\end{enumerate}

Let us take the row of the inverse matrix $s^{-1}$ corresponding to the
trivial representation  $V_0$, then

\be
\sum_{j'\in{\cal S}}(s^{-1})_{{\scs 0}j'} s_{j'j}=\delta_{0,j}
\label{invmat}
\ee
We can consider \eq{invmat} as an analog of the basic integral (\ref{int})
with the obvious action of the co-multiplication:

\be \raisebox{-0.7cm}{
\setlength{\unitlength}{0.0125in}
\begin{picture}(105,55)(0,-10)
\thicklines
\path(25,13)(25,0)
\path(90,39)(90,17)
\path(90,13)(90,0)
\path(79,30)(88,30)
\path(25,40)(25,17)
\path(77,39)(77,17)
\path(77,13)(77,1)
\path(23,30)	(20.060,30.348)
	(17.871,30.464)
	(15.000,30.000)

\path(15,30)	(12.113,27.887)
	(10.000,25.000)

\path(10,25)	(9.613,22.500)
	(10.000,20.000)

\path(10,20)	(12.113,17.113)
	(15.000,15.000)

\path(15,15)	(17.293,14.565)
	(20.080,14.613)
	(25.000,15.000)

\path(25,15)	(29.920,14.613)
	(32.707,14.565)
	(35.000,15.000)

\path(35,15)	(37.887,17.113)
	(40.000,20.000)

\path(40,20)	(40.387,22.500)
	(40.000,25.000)

\path(40,25)	(37.887,27.887)
	(35.000,30.000)

\path(35,30)	(32.000,30.000)

\path(32,30)	(30.266,30.000)
	(27.000,30.000)

\path(75,30)	(70.512,30.407)
	(68.000,30.000)

\path(68,30)	(65.113,27.887)
	(63.000,25.000)

\path(63,25)	(62.613,22.500)
	(63.000,20.000)

\path(63,20)	(65.113,17.113)
	(68.000,15.000)

\path(68,15)	(70.293,14.565)
	(73.080,14.613)
	(78.000,15.000)

\path(78,15)	(81.966,15.000)
	(84.416,15.000)
	(87.000,15.000)
	(89.584,15.000)
	(92.034,15.000)
	(96.000,15.000)

\path(96,15)	(100.000,15.000)

\path(100,15)	(102.887,17.113)
	(105.000,20.000)

\path(105,20)	(105.387,22.500)
	(105.000,25.000)

\path(105,25)	(102.887,27.887)
	(100.000,30.000)

\path(100,30)	(97.129,30.464)
	(94.940,30.348)
	(92.000,30.000)

\put(50,22){\makebox(0,0){$\Big)\! =$}}
\put(0,22){\makebox(0,0){$\Delta\Big($}}
\end{picture}
}
\ee 
Owing to the third condition in
the definition of the modular Hopf algebra, we find the following analog of
the orthogonality of matrix elements

\be
\sum_{i\in{\cal S}}(s^{-1})_{{\scs 0}i}\qtr\Big[
\setlength{\unitlength}{0.0125in}
\raisebox{-0.5in}{
\begin{picture}(42,85)(0,-10)
\thicklines
\path(35,10)(35,30)(5,30)
	(5,10)(35,10)
\path(15,10)(15,0)
\path(25,10)(25,0)
\path(15,30)(15,43)
\path(17,60)(23,60)
\path(25,30)(25,43)
\path(15,70)(15,47)
\path(25,70)(25,47)
\path(13,60)	(10.060,60.348)
	(7.871,60.464)
	(5.000,60.000)

\path(5,60)	(2.113,57.887)
	(0.000,55.000)

\path(0,55)	(-0.387,52.500)
	(0.000,50.000)

\path(0,50)	(2.113,47.113)
	(5.000,45.000)

\path(5,45)	(7.293,44.565)
	(10.080,44.613)
	(15.000,45.000)

\path(15,45)	(18.966,45.000)
	(21.416,45.000)
	(24.000,45.000)
	(26.584,45.000)
	(29.034,45.000)
	(33.000,45.000)

\path(33,45)	(37.000,45.000)

\path(37,45)	(39.887,47.113)
	(42.000,50.000)

\path(42,50)	(42.387,52.500)
	(42.000,55.000)

\path(42,55)	(39.887,57.887)
	(37.000,60.000)

\path(37,60)	(33.000,60.000)

\path(33,60)	(30.572,60.000)
	(27.000,60.000)
\put(20,20){\makebox(0,0){$f$}}
\put(42,65){\makebox(0,0){$\scr i$}}
\put(15,-5){\makebox(0,0){$\scr j_1$}}
\put(25,-5){\makebox(0,0){$\scr j_2$}}
\end{picture}}
\Big]
=\sum_{i\in{\cal S}}(s^{-1})_{{\scs 0}i}\qtr\Big[
\raisebox{-0.65in}{
\setlength{\unitlength}{0.0125in}
\begin{picture}(40,105)(4,-10)
\thicklines
\path(15,90)(20,80)(25,90)
\path(20,80)(20,62)
\path(20,58)(20,50)
\path(15,35)(15,40)(15,45)
	(20,50)(25,45)(25,40)(25,35)
\path(35,15)(35,35)(5,35)
	(5,15)(35,15)
\put(20,23){\makebox(0,0){$f$}}
\path(15,15)(15,5)
\path(25,15)(25,5)
\path(17,75)	(12.900,75.043)
	(9.872,74.906)
	(6.000,74.000)

\path(6,74)	(2.209,71.843)
	(0.000,68.000)

\path(0,68)	(3.099,63.391)
	(8.000,61.000)

\path(8,61)	(12.045,60.083)
	(14.493,59.875)
	(17.059,59.788)
	(22.044,59.849)
	(26.000,60.000)

\path(26,60)	(29.990,60.145)
	(34.000,61.000)

\path(34,61)	(37.798,63.697)
	(40.000,68.000)

\path(40,68)	(38.210,71.682)
	(35.000,74.000)

\path(35,74)	(30.750,75.064)
	(27.453,75.161)
	(23.000,75.000)
\put(42,75){\makebox(0,0){$\scr i$}}
\put(15,0){\makebox(0,0){$\scr j_1$}}
\put(25,0){\makebox(0,0){$\scr j_2$}}
\end{picture}
}\Big]=
\frac{\delta_{j_1,j_2}}{d_{j_1}}
\raisebox{-0.4in}{
\setlength{\unitlength}{0.0125in}
\begin{picture}(30,65)(0,-10)
\thicklines
\path(10,35)	(10.000,38.266)
	(10.000,40.000)

\path(10,40)	(9.806,42.460)
	(10.000,45.000)

\path(10,45)	(11.742,48.161)
	(15.000,50.000)

\path(15,50)	(18.258,48.161)
	(20.000,45.000)

\path(20,45)	(20.194,42.460)
	(20.000,40.000)

\path(20,40)	(20.000,38.266)
	(20.000,35.000)

\path(30,15)(30,35)(0,35)
	(0,15)(30,15)
\path(10,15)	(10.000,11.734)
	(10.000,10.000)

\path(10,10)	(9.806,7.540)
	(10.000,5.000)

\path(10,5)	(11.742,1.839)
	(15.000,0.000)

\path(15,0)	(18.258,1.839)
	(20.000,5.000)

\path(20,5)	(20.194,7.540)
	(20.000,10.000)

\path(20,10)	(20.000,11.734)
	(20.000,15.000)

\put(15,25){\makebox(0,0){$f$}}
\end{picture}
}
\label{orthog}
\ee
for any endomorphism $f$: $V\otimes V\to V\otimes V$. In these formulas,
the $q$-trace is necessary to project out the ideal $I$.

An example of the modular Hopf algebra has been given by Reshetikhin and
Turaev \cite{ReshTur} in the $sl_2$ case. In \rf{TW} the notion of the
quasi-modular Hopf algebra has been introduced by slightly weakening the
irreducibility condition on the modules from $\{V_j\}_{j\in{\cal S}}$. It
still leads to 3-manifold invariants of the Reshetikhin-Turaev type and
therefore sufficient for our purposes as well. Turaev and Wenzl have
constructed examples of quasi-modular Hopf algebras associated with
$\Uq(\mbox{\bf g})$, $q^{\ell}=1$, for all {\bf g} of the $A$, $B$ and $D$
types. 

Thus, we define the qQCD$_3$ partition function at a root of unity in the
same way as in the previous section, using the given above definition of
the integral. One can describe the quantity $Z_{j_1\ldots j_{N_2}}$
appearing in \eq{Zjj} as follows. We consider a Heegaard splitting
$M=H\bigcup_h\wt{H}$. Let us continue a homomorphism $h$ into a small
neighborhood of $\d H$. In other words, $H\bigcap\wt{H}=M^2_g\times[0,1]$,
$\d H=M^2_g\times1$ and $\d\wt{H}=M^2_g\times0$; $\{\alpha\}\in \d H$ and
$\{\gamma\}\in \d \wt{H}$.  Then for any standard embedding of $H$ into
$R^3$, the characteristic curves and the $\alpha$-cycles form a non-trivial
link, ${\cal L}$. They are colored with two sets of representations
$j_1,\ldots,j_{N_2}\in{\cal S}$ and $i_1,\ldots,i_{N_1}\in{\cal S}$. By
using the Reshetikhin-Turaev functor, we calculate the quantum invariant of
the link, $J^{i_1\ldots i_{N_1}}_{j_1\ldots j_{N_2}}({\cal L})$, and sum
over the colors of the $\alpha$-cycles with the weights $(s^{-1})_{{\scs
0}i}$:

\be
Z_{j_1\ldots j_{N_2}}=\sum_{i_1\ldots i_{N_1}\in{\cal S}} 
\prod_{k=1}^{N_1}(s^{-1})_{{\scs 0}i_k} J^{i_1\ldots i_{N_1}}_{j_1\ldots
j_{N_2}}({\cal L})
\ee

\subsection{The topological limit}

If in the root of unity case one chooses the Boltzmann weight coefficients
equal to \mbox{$v_j=(s^{-1})_{0j}$}, one finds the partition function

\be
{\cal Z}_0(C)= \sum_{j_1\ldots j_{N_2}\in{\cal S}} \prod_{k=1}^{N_2}v_{j_k} 
Z_{j_1\ldots j_{N_2}}=\eol\dsty
\sum_{j_1\ldots j_{N_2}\in{\cal S}} \prod_{k=1}^{N_2}v_{j_k} 
\sum_{i_1\ldots i_{N_1}\in{\cal S}} \prod_{k=1}^{N_1}v_{i_k} 
J^{i_1\ldots i_{N_1}}_{j_1\ldots j_{N_2}}({\cal L})
\label{definv}
\ee
where $N_k$ is the number of $k$-dimensional cells in a complex $C$;
$J^{i_1\ldots i_{N_1}}_{j_1\ldots j_{N_2}}({\cal L})$ is the quantum
invariant of a link $\cal L$ given by a Heegaard diagram associated to  the
complex $C$. Let us denote $\omega=\sum_{i\in{\cal S}} v_i d_i$.

\noindent
{\sc theorem~1:} $\I(\M)={\cal Z}_0(C)/\omega^{\scs N_0+N_3-2}$ is a
topological invariant of a manifold $\M$ represented by a complex
$C$. $\I(\M)$ is multiplicative with respect to the
connected sum:

\[
\I(\M)=\I(\M_1)\I(\M_2),\hspace{1pc} \mbox{if}\ \M=\M_1\#\M_2
\]
and $\I(S^3)=1$.

\noindent
{\sc proof} The Heegaard splitting associated to a cell complex $C$ having
$N_k$ cells in the $k$'th dimension gives a handlebody, $H_g$, of the genus
$g=N_1-N_0+1$. Let us fix $g$ independent $\alpha$-cycles of the Heegaard
diagram and take the corresponding integrals (\ie sums over $i$'s in
\eq{definv}) firstly. By applying the CGC decomposition and then using the
orthogonality (\ref{orthog}), we deform the set of the characteristic
curves in the link $\cal L$ into some 3-valent ribbon graph $\cal G$. Every
application of \eq{orthog} distroys a handle of $H_g$. Therefore, having
taken the $g$ integrals, we obtain a spherical ribbon 3-valent graph $\cal
G$ plus a collection of $N_1-g=N_0-1$ disjoint unlinked loops corresponding
to the rest of the $\alpha$-cycles. The integrals associated to them give
$\omega^{\scs N_0-1}$. Now, we can recover the intial configuration of the
characteristic curves\footnote{\mbox{}$^)$ or another one equivalent to it.} 
by restoring the $g$ integrals corresponding to the
independent $\alpha$-cycles. In this way we obtain a cell decomposition of
$\M$ with only one 0-cell and every 1-cell corresponding to a generator
of $\pi_1(\M)$. This procedure is the direct analog of fixing an axial
gauge in LGT.

The Heegaard splitting is obviously symmetric with respect to the
Poincar\'{e} duality, therefore we can repeat the previous procedure with
roles of the $\alpha$-cycles and the characteristic curves interchanged. In
this way we fix a set of $g$ independent characteristic curves and pick up
the factor $\omega^{\scs N_3-1}$. Thus, we finish with some balanced
presentation of $\pi_1(\M)$.

To prove the topological invariance, we have to show  that $\I(\M)$ is not
changed [i]~by Dehn twists on contractible loops, [ii]~by the cycle slide
and [iii]~by the stabilization.

\noindent
i)~Invariance under the Dehn twists on loops contractible inside $H_g$ is
obvious. By taking the $g$ integrals, we cut all handles and always get the
same 3-valent ribbon graph $\cal G$.\\
ii)~Invariance under the cycle slide follows from the analog of Haar
measure invariance as illustrated in Figure~2.

\begin{figure}
\[
\sum v_i \sum v_j\; \qtr\Bigg[
\raisebox{-0.6in}{
\setlength{\unitlength}{0.0125in}
\begin{picture}(108,125)(0,-10)
\thicklines
\path(25,110)(25,84)
\path(35,110)(35,84)
\path(15,77)(15,58)
\path(25,77)(25,59)
\path(35,77)(35,59)
\path(65,110)(65,58)
\path(75,110)(75,59)
\path(85,110)(85,59)
\path(37,65)(62,65)
\path(25,52)(25,40)
\path(15,52)(15,40)
\path(35,52)(35,40)
\path(65,51)(65,40)
\path(75,52)(75,40)
\path(85,52)(85,40)
\path(95,10)(95,40)(5,40)
	(5,10)(95,10)
\path(15,110)(15,84)
\path(15,10)(15,0)
\path(25,10)(25,0)
\path(35,10)(35,0)
\path(65,10)(65,0)
\path(75,10)(75,0)
\path(85,10)(85,0)
\path(17,90)(22,90)
\path(28,90)(33,90)
\path(18,65)(23,65)
\path(28,65)(33,65)
\path(67,65)(72,65)
\path(77,65)(82,65)
\path(11,90)	(7.153,90.348)
	(5.000,90.000)

\path(5,90)	(1.839,88.257)
	(0.000,85.000)

\path(0,85)	(1.839,81.743)
	(5.000,80.000)

\path(5,80)	(7.540,79.807)
	(10.000,80.000)

\path(10,80)	(12.971,80.000)
	(16.610,80.000)
	(20.693,80.000)
	(25.000,80.000)
	(29.307,80.000)
	(33.390,80.000)
	(37.029,80.000)
	(40.000,80.000)

\path(40,80)	(42.460,79.807)
	(45.000,80.000)

\path(45,80)	(48.161,81.743)
	(50.000,85.000)

\path(50,85)	(48.161,88.257)
	(45.000,90.000)

\path(45,90)	(42.488,90.407)
	(38.000,90.000)

\path(11,65)	(7.153,65.348)
	(5.000,65.000)

\path(5,65)	(1.839,63.257)
	(0.000,60.000)

\path(0,60)	(1.839,56.743)
	(5.000,55.000)

\path(5,55)	(7.540,54.807)
	(10.000,55.000)

\path(10,55)	(13.961,55.000)
	(18.813,55.000)
	(21.480,55.000)
	(24.258,55.000)
	(27.110,55.000)
	(30.000,55.000)
	(32.890,55.000)
	(35.742,55.000)
	(38.520,55.000)
	(41.187,55.000)
	(46.039,55.000)
	(50.000,55.000)

\path(50,55)	(53.961,55.000)
	(58.813,55.000)
	(61.480,55.000)
	(64.258,55.000)
	(67.110,55.000)
	(70.000,55.000)
	(72.890,55.000)
	(75.742,55.000)
	(78.520,55.000)
	(81.187,55.000)
	(86.039,55.000)
	(90.000,55.000)

\path(90,55)	(92.460,54.807)
	(95.000,55.000)

\path(95,55)	(98.161,56.743)
	(100.000,60.000)

\path(100,60)	(98.161,63.257)
	(95.000,65.000)

\path(95,65)	(92.488,65.407)
	(88.000,65.000)
\put(102,67){\makebox(0,0){$\scs i$}}
\put(50,25){\makebox(0,0){$f$}}
\put(51,91){\makebox(0,0){$\scs j$}}
\end{picture}
}\Bigg]\mbox{\Large =}
\sum v_i \sum v_j\; \qtr\Bigg[
\raisebox{-0.6in}{
\setlength{\unitlength}{0.0125in}
\begin{picture}(103,135)(0,-10)
\thicklines
\path(71,70)(71,45)
\path(81,70)(81,45)
\path(91,15)(91,45)(1,45)
	(1,15)(91,15)
\path(61,15)(61,0)
\path(71,15)(71,0)
\path(81,15)(81,0)
\path(11,15)(11,0)
\path(21,15)(21,0)
\path(31,15)(31,0)
\path(61,115)(61,74)
\path(71,115)(71,74)
\path(81,115)(81,74)
\path(24,64)(24,70)
\path(61,70)(61,45)
\path(27,82)(57,82)
\path(24,100)(24,87)
\path(24,83)(24,75)
\path(64,82)(69,82)
\path(74,82)(79,82)
\path(11,45)	(10.710,48.206)
	(11.000,50.000)

\path(11,50)	(12.742,53.161)
	(16.000,55.000)

\path(16,55)	(19.258,53.161)
	(21.000,50.000)

\path(21,50)	(21.290,48.206)
	(21.000,45.000)

\path(16,55)	(15.710,58.206)
	(16.000,60.000)

\path(16,60)	(18.113,62.887)
	(21.000,65.000)

\path(21,65)	(23.500,65.387)
	(26.000,65.000)

\path(26,65)	(28.887,62.887)
	(31.000,60.000)

\path(31,60)	(31.194,57.460)
	(31.000,55.000)

\path(31,55)	(31.000,51.531)
	(31.000,48.765)
	(31.000,45.000)

\path(26,95)	(29.206,95.290)
	(31.000,95.000)

\path(31,95)	(34.161,93.257)
	(36.000,90.000)

\path(36,90)	(34.161,86.743)
	(31.000,85.000)

\path(31,85)	(28.460,84.807)
	(26.000,85.000)

\path(26,85)	(23.500,85.000)
	(21.000,85.000)

\path(21,85)	(18.540,84.807)
	(16.000,85.000)

\path(16,85)	(12.839,86.743)
	(11.000,90.000)

\path(11,90)	(12.839,93.257)
	(16.000,95.000)

\path(16,95)	(17.794,95.290)
	(21.000,95.000)

\path(11,120)	(10.710,116.794)
	(11.000,115.000)

\path(11,115)	(12.742,111.839)
	(16.000,110.000)

\path(16,110)	(19.258,111.839)
	(21.000,115.000)

\path(21,115)	(21.290,116.794)
	(21.000,120.000)

\path(16,110)	(15.710,106.794)
	(16.000,105.000)

\path(16,105)	(18.113,102.113)
	(21.000,100.000)

\path(21,100)	(23.500,99.613)
	(26.000,100.000)

\path(26,100)	(28.887,102.113)
	(31.000,105.000)

\path(31,105)	(31.194,107.540)
	(31.000,110.000)

\path(31,110)	(31.000,113.469)
	(31.000,116.235)
	(31.000,120.000)

\path(20,82)	(16.235,82.000)
	(13.469,82.000)
	(10.000,82.000)

\path(10,82)	(7.540,82.193)
	(5.000,82.000)

\path(5,82)	(1.839,80.257)
	(0.000,77.000)

\path(0,77)	(1.839,73.743)
	(5.000,72.000)

\path(5,72)	(7.293,71.565)
	(10.080,71.613)
	(15.000,72.000)

\path(15,72)	(18.466,72.000)
	(22.711,72.000)
	(27.476,72.000)
	(29.972,72.000)
	(32.500,72.000)
	(35.028,72.000)
	(37.524,72.000)
	(42.289,72.000)
	(46.534,72.000)
	(50.000,72.000)

\path(50,72)	(50.000,72.000)

\path(50,72)	(53.466,72.000)
	(57.711,72.000)
	(62.476,72.000)
	(64.972,72.000)
	(67.500,72.000)
	(70.028,72.000)
	(72.524,72.000)
	(77.289,72.000)
	(81.534,72.000)
	(85.000,72.000)

\path(85,72)	(87.460,71.807)
	(90.000,72.000)

\path(90,72)	(93.161,73.743)
	(95.000,77.000)

\path(95,77)	(93.161,80.257)
	(90.000,82.000)

\path(90,82)	(87.488,82.407)
	(83.000,82.000)
\put(97,84){\makebox(0,0){$\scs i$}}
\put(46,30){\makebox(0,0){$f$}}
\put(39,94){\makebox(0,0){$\scs j$}}
\end{picture}
}\Bigg]\mbox{\Large =}
\]
\[
\frac1d\sum v_i\; \qtr\Bigg[
\raisebox{-0.6in}{
\setlength{\unitlength}{0.0125in}
\begin{picture}(103,130)(0,-10)
\thicklines
\path(70,115)(70,79)
\path(80,115)(80,79)
\path(60,71)(60,60)
\path(70,72)(70,60)
\path(80,72)(80,60)
\path(90,30)(90,60)(0,60)
	(0,30)(90,30)
\path(60,30)(60,10)
\path(70,30)(70,10)
\path(60,115)(60,78)
\path(80,30)(80,10)
\path(62,85)(67,85)
\path(72,85)(77,85)
\path(56,85)	(52.153,85.348)
	(50.000,85.000)

\path(50,85)	(46.839,83.257)
	(45.000,80.000)

\path(45,80)	(46.839,76.743)
	(50.000,75.000)

\path(50,75)	(50.000,75.000)

\path(50,75)	(53.466,75.000)
	(57.711,75.000)
	(62.476,75.000)
	(64.972,75.000)
	(67.500,75.000)
	(70.028,75.000)
	(72.524,75.000)
	(77.289,75.000)
	(81.534,75.000)
	(85.000,75.000)

\path(85,75)	(87.460,74.807)
	(90.000,75.000)

\path(90,75)	(93.161,76.743)
	(95.000,80.000)

\path(95,80)	(93.161,83.257)
	(90.000,85.000)

\path(90,85)	(87.488,85.407)
	(83.000,85.000)

\path(15,15)	(15.000,11.734)
	(15.000,10.000)

\path(15,10)	(14.806,7.540)
	(15.000,5.000)

\path(15,5)	(17.113,2.113)
	(20.000,0.000)

\path(20,0)	(22.500,-0.387)
	(25.000,0.000)

\path(25,0)	(27.887,2.113)
	(30.000,5.000)

\path(30,5)	(30.194,7.540)
	(30.000,10.000)

\path(30,10)	(30.000,12.203)
	(30.000,15.000)
	(30.000,17.797)
	(30.000,20.000)

\path(30,20)	(30.000,23.469)
	(30.000,26.235)
	(30.000,30.000)

\path(10,30)	(10.000,26.734)
	(10.000,25.000)

\path(10,25)	(9.806,22.540)
	(10.000,20.000)

\path(10,20)	(11.742,16.839)
	(15.000,15.000)

\path(15,15)	(18.258,16.839)
	(20.000,20.000)

\path(20,20)	(20.194,22.540)
	(20.000,25.000)

\path(20,25)	(20.000,26.734)
	(20.000,30.000)

\path(10,60)	(10.000,63.266)
	(10.000,65.000)

\path(10,65)	(9.806,67.460)
	(10.000,70.000)

\path(10,70)	(11.742,73.161)
	(15.000,75.000)

\path(15,75)	(18.258,73.161)
	(20.000,70.000)

\path(20,70)	(20.194,67.460)
	(20.000,65.000)

\path(20,65)	(20.000,63.266)
	(20.000,60.000)

\path(15,75)	(15.000,78.266)
	(15.000,80.000)

\path(15,80)	(14.806,82.460)
	(15.000,85.000)

\path(15,85)	(17.113,87.887)
	(20.000,90.000)

\path(20,90)	(22.500,90.387)
	(25.000,90.000)

\path(25,90)	(27.887,87.887)
	(30.000,85.000)

\path(30,85)	(30.194,82.460)
	(30.000,80.000)

\path(30,80)	(30.000,77.797)
	(30.000,75.000)
	(30.000,72.203)
	(30.000,70.000)

\path(30,70)	(30.000,66.531)
	(30.000,63.765)
	(30.000,60.000)
\put(97,87){\makebox(0,0){$\scs i$}}
\put(45,45){\makebox(0,0){$f$}}
\end{picture}
}\Bigg]\mbox{\Large =}
\sum v_i \sum v_j\; \qtr\Bigg[
\raisebox{-0.6in}{
\setlength{\unitlength}{0.0125in}
\begin{picture}(108,125)(0,-10)
\thicklines
\path(25,110)(25,84)
\path(35,110)(35,84)
\path(15,77)(15,40)
\path(25,77)(25,40)
\path(35,77)(35,40)
\path(65,110)(65,58)
\path(75,110)(75,59)
\path(85,110)(85,59)
\path(65,51)(65,40)
\path(75,52)(75,40)
\path(85,52)(85,40)
\path(95,10)(95,40)(5,40)
	(5,10)(95,10)
\path(15,10)(15,0)
\path(15,110)(15,84)
\path(25,10)(25,0)
\path(35,10)(35,0)
\path(65,10)(65,0)
\path(75,10)(75,0)
\path(85,10)(85,0)
\path(17,90)(22,90)
\path(28,90)(33,90)
\path(67,65)(72,65)
\path(77,65)(82,65)
\path(11,90)	(7.153,90.348)
	(5.000,90.000)

\path(5,90)	(1.839,88.257)
	(0.000,85.000)

\path(0,85)	(1.839,81.743)
	(5.000,80.000)

\path(5,80)	(7.540,79.807)
	(10.000,80.000)

\path(10,80)	(12.971,80.000)
	(16.610,80.000)
	(20.693,80.000)
	(25.000,80.000)
	(29.307,80.000)
	(33.390,80.000)
	(37.029,80.000)
	(40.000,80.000)

\path(40,80)	(42.460,79.807)
	(45.000,80.000)

\path(45,80)	(48.161,81.743)
	(50.000,85.000)

\path(50,85)	(48.161,88.257)
	(45.000,90.000)

\path(45,90)	(42.488,90.407)
	(38.000,90.000)

\path(61,65)	(57.153,65.348)
	(55.000,65.000)

\path(55,65)	(51.839,63.257)
	(50.000,60.000)

\path(50,60)	(51.839,56.743)
	(55.000,55.000)

\path(55,55)	(55.000,55.000)

\path(55,55)	(58.466,55.000)
	(62.711,55.000)
	(67.476,55.000)
	(69.972,55.000)
	(72.500,55.000)
	(75.028,55.000)
	(77.524,55.000)
	(82.289,55.000)
	(86.534,55.000)
	(90.000,55.000)

\path(90,55)	(92.460,54.807)
	(95.000,55.000)

\path(95,55)	(98.161,56.743)
	(100.000,60.000)

\path(100,60)	(98.161,63.257)
	(95.000,65.000)

\path(95,65)	(92.488,65.407)
	(88.000,65.000)
\put(102,67){\makebox(0,0){$\scs i$}}
\put(50,25){\makebox(0,0){$f$}}
\put(51,91){\makebox(0,0){$\scs j$}}
\end{picture}
}\Bigg]
\]
\caption{Invariance under the cycle slide.}
\end{figure}

\noindent
iii)~The stabilization consists in adding a handle to $H_g$ and extending a
gluing homomorphism $h$ by the identity on its boundary. It amounts to the
addition to $\cal L$ of one $\alpha$-cycle and one characteristic curve
forming the Hopf link. Therefore, the integration associated to the new
handle attaches the trivial representation to the new characteristic curve
and it is unimportant how it is linked with the other $\alpha$-cycles.

To show the multiplicative nature of the invariant, let us choose such a
cell decomposition of $\M=\M_1\#\M_2$ that a sphere dividing $\M_1$ and
$\M_2$ consists of only one 0-cell and one 2-cell. Then the corresponding
characteristic curve is linked with no $\alpha$-cycle and the corresponding
link $\cal L$ in \eq{definv} has two connected components.

The normalization $\I(S^3)=1$ follows from the observation that the Hopf
link corresponds to a genus 1 Heegaard splitting of the sphere. $\Box$

\noindent
{\em Remarks}:~1)~The meaning of the choice of the Boltzmann weight
coefficients made in \eq{definv} is clear. They corresponds to the
$\delta$-function weights. Therefore, ${\cal Z}_0(\M)$ can be regarded as a
generalization of \eq{fingr}. In contrast to the finite group partition
function, the $q$-deformed model is obviously self-dual with respect to the
Poincar\'{e} duality of 3-folds.\\ 
2)~Let us consider a simplicial complex $C^{(s)}$. If we take in the 
expression (\ref{definv}) for the partition function ${\cal Z}_0(C^{(s)})$ 
all the sums associated to triangles in $C^{(s)}$ prior to the others, 
then the answer is identical to the definition of the
Turaev-Viro invariant. Indeed, for each triangle we find the tangle
equivalent to the product of two $3j$-symbols:

\be \sum v_i \raisebox{-0.45in}{ 
\setlength{\unitlength}{0.0125in}
\begin{picture}(60,70)(0,-10)
\thicklines
\path(30,55)(30,24)
\path(40,55)(40,24)
\path(20,17)(20,0)
\path(20,55)(20,24)
\path(30,17)(30,0)
\path(17,40)	(12.512,40.407)
	(10.000,40.000)

\path(10,40)	(6.951,38.595)
	(3.679,36.515)
	(1.067,33.677)
	(0.000,30.000)

\path(0,30)	(1.067,26.322)
	(3.679,23.485)
	(6.951,21.405)
	(10.000,20.000)

\path(10,20)	(12.540,19.807)
	(15.000,20.000)

\path(15,20)	(17.476,20.000)
	(20.508,20.000)
	(23.911,20.000)
	(27.500,20.000)
	(31.089,20.000)
	(34.492,20.000)
	(37.524,20.000)
	(40.000,20.000)

\path(40,20)	(44.920,19.613)
	(47.707,19.565)
	(50.000,20.000)

\path(50,20)	(53.049,21.405)
	(56.321,23.485)
	(58.933,26.322)
	(60.000,30.000)

\path(60,30)	(58.933,33.678)
	(56.321,36.515)
	(53.049,38.595)
	(50.000,40.000)

\path(50,40)	(47.488,40.407)
	(43.000,40.000)

\path(40,17)(40,0)
\path(22,40)(28,40)
\path(32,40)(38,40)
\put(20,-5){\makebox(0,0){$\scs j_1$}}
\put(30,-5){\makebox(0,0){$\scs j_2$}}
\put(40,-5){\makebox(0,0){$\scs j_3$}}
\end{picture}
}= \sum v_i \raisebox{-0.45in}{ 
\setlength{\unitlength}{0.0125in}
\begin{picture}(26,75)(0,-10)
\thicklines
\path(13,20)(13,25)
\path(5,50)	(4.710,46.794)
	(5.000,45.000)

\path(5,45)	(7.113,42.113)
	(10.000,40.000)

\path(10,40)	(12.500,39.613)
	(15.000,40.000)

\path(15,40)	(17.887,42.113)
	(20.000,45.000)

\path(20,45)	(20.194,47.540)
	(20.000,50.000)

\path(20,50)	(20.000,53.469)
	(20.000,56.235)
	(20.000,60.000)

\path(13,40)(13,28)
\path(0,60)	(-0.290,56.794)
	(0.000,55.000)

\path(0,55)	(1.743,51.839)
	(5.000,50.000)

\path(5,50)	(8.258,51.839)
	(10.000,55.000)

\path(10,55)	(10.290,56.794)
	(10.000,60.000)

\path(11,36)	(7.794,36.290)
	(6.000,36.000)

\path(6,36)	(2.839,34.257)
	(1.000,31.000)

\path(1,31)	(2.839,27.743)
	(6.000,26.000)

\path(6,26)	(8.540,25.807)
	(11.000,26.000)

\path(11,26)	(15.920,25.613)
	(18.707,25.565)
	(21.000,26.000)

\path(21,26)	(24.161,27.743)
	(26.000,31.000)

\path(26,31)	(24.161,34.257)
	(21.000,36.000)

\path(21,36)	(19.206,36.290)
	(16.000,36.000)

\path(5,10)	(4.710,13.206)
	(5.000,15.000)

\path(5,15)	(7.113,17.887)
	(10.000,20.000)

\path(10,20)	(12.500,20.387)
	(15.000,20.000)

\path(15,20)	(17.887,17.887)
	(20.000,15.000)

\path(20,15)	(20.194,12.460)
	(20.000,10.000)

\path(20,10)	(20.000,6.531)
	(20.000,3.765)
	(20.000,0.000)

\path(0,0)	(-0.290,3.206)
	(0.000,5.000)

\path(0,5)	(1.743,8.161)
	(5.000,10.000)

\path(5,10)	(8.258,8.161)
	(10.000,5.000)

\path(10,5)	(10.290,3.206)
	(10.000,0.000)
\put(0,-5){\makebox(0,0){$\scs j_1$}}
\put(10,-5){\makebox(0,0){$\scs j_2$}}
\put(20,-5){\makebox(0,0){$\scs j_3$}}
\end{picture}
}=
\frac1{d_{j_3}} \raisebox{-0.4in}{ 
\setlength{\unitlength}{0.0125in}
\begin{picture}(20,70)(0,-10)
\thicklines
\path(0,55)	(-0.290,51.794)
	(0.000,50.000)

\path(0,50)	(1.743,46.839)
	(5.000,45.000)

\path(5,45)	(8.258,46.839)
	(10.000,50.000)

\path(10,50)	(10.290,51.794)
	(10.000,55.000)

\path(5,45)	(4.710,41.794)
	(5.000,40.000)

\path(5,40)	(7.113,37.113)
	(10.000,35.000)

\path(10,35)	(12.500,34.613)
	(15.000,35.000)

\path(15,35)	(17.887,37.113)
	(20.000,40.000)

\path(20,40)	(20.194,42.540)
	(20.000,45.000)

\path(20,45)	(20.000,48.469)
	(20.000,51.235)
	(20.000,55.000)

\path(5,10)	(4.710,13.206)
	(5.000,15.000)

\path(5,15)	(7.113,17.887)
	(10.000,20.000)

\path(10,20)	(12.500,20.387)
	(15.000,20.000)

\path(15,20)	(17.887,17.887)
	(20.000,15.000)

\path(20,15)	(20.194,12.460)
	(20.000,10.000)

\path(20,10)	(20.000,6.531)
	(20.000,3.765)
	(20.000,0.000)

\path(0,0)	(-0.290,3.206)
	(0.000,5.000)

\path(0,5)	(1.743,8.161)
	(5.000,10.000)

\path(5,10)	(8.258,8.161)
	(10.000,5.000)

\path(10,5)	(10.290,3.206)
	(10.000,0.000)
\put(0,-5){\makebox(0,0){$\scs j_1$}}
\put(10,-5){\makebox(0,0){$\scs j_2$}}
\put(20,-5){\makebox(0,0){$\scs j_3$}}
\end{picture}
} 
\ee 
By closing all the tangles, we obtain
a Racah-Wiegner $6j$-symbol

\be
\left\{\ba{ccc}j_1&j_2&j_3\\j_4&j_5&j_6\ea\right\} =
\frac1{d_{j_6}\sqrt{d_{j_2}d_{j_5}}}
\raisebox{-2cm}{\bp(40,40)(-5,-5)
\thicklines
\put(5,15){\oval(10,10)[b]}
\put(12.5,10){\oval(15,10)[b]}
\put(17.75,5){\oval(14.5,10)[b]}
\put(7.5,20){\oval(15,10)[t]}
\put(15,15){\oval(10,10)[t]}
\put(16.25,25){\oval(17.5,10)[t]}
\put(1.5,15){\line(0,1){5}}
\put(21.5,10){\line(0,1){5}}
\put(26.5,5){\line(0,1){20}}
\put(1,26){$\scr j_1$}
\put(8,18){$\scr j_3$}
\put(15,24){$\scr j_2$}
\put(16,8){$\scr j_4$}
\put(5,3){$\scr j_5$}
\put(27,15){$\scr j_6$}
\ep}
\label{6j}
\ee
inside each tetrahedron of the simplicial complex. The indices, 
$j_1,\ldots,j_{\scs N_1}$, are
attached to  1-simplexes of $C^{(s)}$. Taking the sums over them we arrive
at the Turaev-Viro state sum invariant \cite{TV}:
\be
\I_0(C^{(s)}) = \omega^{\scs N_1-N_2+2} 
\sum_{\{j_k\in{\cal S}\}} \prod_{k=1}^{N_1} v_{j_k}\,
\prod_{t=1}^{N_3}
\left\{\ba{ccc}j_{t_1}&j_{t_2}&j_{t_3}\\j_{t_4}&j_{t_5}&j_{t_6}\ea\right\}
\label{TuV}
\ee 
where the 6-tuple $(t_1,\ldots,t_6)$ denotes six edges of the $t$-th
tetrahedron. Explicit expressions for $(s^{-1})_{ij}$ in the $sl_2$ case
are given in \rf{ReshTur}. Thus, \eq{definv} can be regarded as a general
definition of the Turaev-Viro invariant.\\ 3)~The expression for ${\cal
Z}_0(\M)$ given in \eq{definv} coincides with the Reshetikhin-Turaev
construction of 3-fold invariants $I_{RT}$ via the surgery representation
\cite{ReshTur}.  Therefore, ${\cal Z}_0(\M)$ is automatically invariant
under the Kirby calculus applied formally to the link $\cal L$. It means
that, given a manifold $\M$, there exists another one $\cal N$ such that
$\I(\M)=I_{RT}({\cal N})$. As $\I(\M)=|I_{RT}(\M)|^2$, we conclude that
${\cal N}\cong \M\#\overline{\M}$ ($\overline{\M}$ is $\M$ with the
opposite orientation). A simple illustration in the case of lens spaces can
be found in \rf{B2}.

\subsection{Bounded manifolds and links}

Every set of disjoint simple closed curves $\{\gamma\}$ on a handlebody $H$
determines a bounded 3-manifold $\M$ constructed by gluing plates to annular
neighborhoods of the curves. It can be shown that every orientable bounded
3-manifold can be obtained in this way. The handlebody in this construction
is a tubular neighborhood of a 1-skeleton of $\M$. Therefore we can
straightforwardly apply the qQCD$_3$ functor in the bounded case. For it,
we [i]~fix a system of $\alpha$-cycles on $\d H$; [ii]~color curves from
$\{\gamma\}$ with $\Uq$ irreps; [iii]~repeat steps 3, 4 and 5 from
Definition~1 without any modification. It suggest the following
interpretation of our construction. A {\em spine} is a 2-dimensional
polyhedron which can be embedded in some 3-manifold. Any 3-manifold with a
boundary collapses to a spine. Let us delete a ball from every 3-cell of a
closed complex $C$. In such a way we obtain a bounded manifold which
collapses to a 2-skeleton $K_2$ of $C$. If $C$ is dual to a simplicial
complex, $K_2$ is called a {\em standard spine}. Matveev has introduced two
moves which relate all standard spines of the same manifold \cite{Mat}. It
can be easily shown that $\I(\M)$ from the previous section is invariant
under the Matveev moves.

The definition of the qQCD$_3$ functor uses an immersion of $K_2$ into
$R^3$ and depends on it. It seems to be an intrinsic feature of q-deformed
LGT rather than a defect of our presentation. Only gauge invariant
singlet quantities (the partition function, for example) are independent of
a way $K_2$ is immersed into $R^3$.

One of the advantages of our presentation of qQCD$_3$ is a relative
simplicity of introducing Wilson loops in it. In classical LGT, a loop
average is defined as

\be
A(L_1,\ldots,L_m)=\frac1{{\cal Z}_{\beta}}\int_G\prod_{\ell} dg_{\ell}\: 
\prod_f W_\beta(h_{\d f}) 
\prod_{i=1}^m \tr_{\mbox{}_{V_{j_i}}}[h_{{\scs L}_i}]
\label{wloop}
\ee
where $\{L\}$ are $m$ closed curves embedded into a 1-skeleton $K^1$ of a
complex $C$. We color the $i$-th curve with a representation $j_i$ of a
gauge group $G$. The holonomy $h_{\scs L}$ is defined in \eq{arg}.

In the $q$-deformed case, we have to specify a link formed by the
collection of curves in a manifold $\M\cong C$. For it we represent the
curves $\{L\}$ by a set of disjoint ribbon loops on a boundary $\d H'$ of a
handlebody $H'\subset H$ (as usual, $H'$ and $H$ are tubular neigborhoods
of $K^1$ and $H'$ lies inside $H$: $H\bigcap H'=H'$, $H\bigcup H'=H$). If
it is not possible, then one has to take a finer subdivision of $\M$. 
In the case of links in $R^3$, it is a standard technical trick to realize
a link as a system of disjoint loops on a handlebody embedded into
$R^3$. And we simply use it as a definition.

We can apply the qQCD$_3$ functor to such a composite handlebody without
any additional modification. Loops from $\{L\}$ enters on equal footing
with characteristic curves. One can repeat the same argument as in the
partition function case to prove that an answer is independent of an
embedding of $H$ in $R^3$.  However, it does not mean that the
$q$-deformation of \eq{wloop} gives no non-trivial knot invariant.  Let us
consider a link in $R^3$. There has to exist a trivial embedding such that
characteristic curves of a Heegaard diagram lying on $\d H$ are unlinked
and contractible in $R^3\setminus H$. Therefore the sums over their colors
disjoin $\alpha$-cycles on $\d H$ and the link of curves $\{L\}$ on $\d H'$
(in other words, cut handles of the complementary handlebody $S^3\setminus
H$).  What remains is exactly the Jones polynomial associated to the link
$\{L\}$. 

The comprehensive treatment of quantum invariants of links and
3-valent graphs in 3-manifolds can be found in \rf{T}.

\newsection{qQCD$_2$}

We define qQCD$_2$ functor by applying the qQCD$_3$ one to an embedding of
an oriented 2-manifold $M^2_g$ in $R^3$. In the topological limit, we can
consider the simplest cell decomposition of $M^2_g$ consisting of a single
2-cell and $2g$ 1-cells.  A tubular neigborhood of its 1-skeleton is a
handlebody $H$ of the genus $2g$. The Heegaard diagram has only one
characteristic curve. Each integral destroys a handle of $H$ and
contributes a factor $1/d_j$ to an answer. The calculation is reduced to a
repeated application of \eq{orthog} and one easily gets

\be
\I(M^2_g)=\omega^{2g-1}{\cal Z}_0(M^2_g)=
\omega^{2g-1}\sum_{j\in{\cal S}}v_j Z_j=\sum_{j\in{\cal S}} v_j 
\Bigg(\frac{d_j}{\omega}\Bigg)^{1-2g}
\ee

Let us consider a concrete example of the quantum group $\Uq(sl(2))$, at
$q=e^{\frac{2\pi i}{k+2}}$. In this case, the set of modules in the
definition of the modular Hopf algebra is given by the fusion ring $V_j$
($j=0,\frac12,1,\frac32,\ldots,\frac{k}2$) and 
$d_j=\frac{\sin(\frac{2j+1}{k+2}\pi)}{\sin\frac{\pi}{k+2}}$;
$\omega=\frac{\sqrt{(k+2)/2}}{\sin\frac{\pi}{k+2}}$; $v_j=d_j/\omega$. 
We find 

\be
\I(M^2_g)=\sum_{j=0,\frac12,\ldots}^{k/2}
\Bigg(\frac{2\sin^2\Big(\frac{2j+1}{k+2}\pi\Big)}{k+2}\Bigg)^{1-g}
\ee
These are known as Verlinde's numbers. They are all integer and equal to
the dimensions of spaces of conformal blocks in the WZW model on a genus
$g$ Riemann surface. 

If one starts with a more complicated cell decomposition of a Riemann
surface, then one has simply apply the orthogonality relation
(\ref{orthog}) till all handles of $H$ are destroyed. 

In two dimensions, local properties of qQCD$_2$ functor can be formalized
in a pure algebraic way. For it, let us cut from $M^2_g$ a piece which can
be projected on a plane $R^2$. It gives a subdivision (triangulation, say)
of some region on the plane. There is a natural cyclic order of edges
incident to a vertex. Following Fock and Rosly \cite{FR}, one introduces a
{\em ciliation} at every vertex, \ie breaks this order. Let us say that an
edge $\ell_1$ goes after $\ell_2$ ($\ell_1>\ell_2$), if an anti-clockwise
angle $\varphi(\ell_1)$ between the edge $\ell_1$ and the $x$-axis is bigger
than an angle $\varphi(\ell_2)$ between $\ell_2$ and the $x$-axis:
$\varphi(\ell_1)>\varphi(\ell_2)$. We assume that no edge is parallel to the
$x$-axis, and orient edges in the $y$-direction. Say, put an arrow at an
end having a bigger $y$ coordinate. Assuming that any two vertices are
connected at most by one edge, we can numerate edges by
ordered pairs of vertices 
$(i,j)\cong \mbox{}_{\mbox{}_{\scr j}}\!\!\nearrow^{\mbox{}^{\scr i}}$.  

Alekseev, Grosse and Schomerus have introduced the following algebra of
gauge fields $U_{(i,j)}$ \cite{AGS}:

\begin{description}
\item[i)] If two edges have no common vertices, fields are co-commutative:
\[
\stackrel{1}{U}_{(i,j)}\stackrel{2}{U}_{(n,m)}=
\stackrel{2}{U}_{(n,m)}\stackrel{1}{U}_{(i,j)}
\]
here $i,j,n$ and $m$ are all distinct.
\item[ii)] If two edges share a vertex, then
\[
\stackrel{1}{U}_{(i,j)}\stackrel{2}{U}_{(k,j)}=\left\{\ba{ll}
\stackrel{2}{U}_{(k,j)}\stackrel{1}{U}_{(i,j)}\r_{12} & 
\mbox{if $(i,j)>(k,j)$}\\
\stackrel{2}{U}_{(k,j)}\stackrel{1}{U}_{(i,j)}\r^{-1}_{12} & 
\mbox{if $(i,j)<(k,j)$}
\ea\right.
\]
We picture these relations as
\[
\raisebox{-0.6in}{
\setlength{\unitlength}{0.0125in}
\begin{picture}(130,106)(0,-10)
\thicklines
\path(50,35)(50,55)(30,55)
	(30,35)(50,35)
\dottedline{5}(10,75)(10,55)
\dottedline{5}(40,75)(40,55)
\path(100,35)(100,55)(80,55)
	(80,35)(100,35)
\path(130,35)(130,55)(110,55)
	(110,35)(130,35)
\path(10,35)(10,15)
\path(40,35)(40,15)
\dottedline{5}(120,55)(120,60)(120,65)
	(90,80)(90,85)(90,90)
\path(20,35)(20,55)(0,55)
	(0,35)(20,35)
\dottedline{5}(120,91)(120,86)(120,81)(110,75)
\dottedline{5}(90,55)(90,60)(90,65)(102,71)
\path(120,35)	(120.000,31.734)
	(120.000,30.000)

\path(120,30)	(120.306,27.567)
	(120.000,25.000)

\path(120,25)	(117.366,22.179)
	(113.775,20.169)
	(109.547,18.699)
	(105.000,17.500)
	(100.453,16.301)
	(96.225,14.831)
	(92.634,12.821)
	(90.000,10.000)

\path(90,10)	(89.694,7.433)
	(90.000,5.000)

\path(90,5)	(90.000,3.266)
	(90.000,0.000)

\path(120,0)	(120.000,3.266)
	(120.000,5.000)

\path(120,5)	(120.276,7.439)
	(120.000,10.000)

\path(120,10)	(116.846,13.020)
	(114.002,14.445)
	(110.000,16.000)

\path(90,35)	(90.000,31.734)
	(90.000,30.000)

\path(90,30)	(89.694,27.567)
	(90.000,25.000)

\path(90,25)	(93.855,21.698)
	(97.251,20.344)
	(102.000,19.000)

\put(65,43){\makebox(0,0){\Large =}}
\put(10,45){\makebox(0,0){$a$}}
\put(40,45){\makebox(0,0){$b$}}
\put(90,45){\makebox(0,0){$b$}}
\put(120,45){\makebox(0,0){$a$}}
\put(25,0){\makebox(0,0){$j$}}
\put(5,85){\makebox(0,0){$i$}}
\put(45,85){\makebox(0,0){$k$}}
\put(105,5){\makebox(0,0){$j$}}
\put(85,90){\makebox(0,0){$i$}}
\put(125,90){\makebox(0,0){$k$}}
\end{picture}
}\hspace{1in}\mbox{if $(i,j)>(k,j)$}
\]
\[
\raisebox{-0.7in}{
\setlength{\unitlength}{0.0125in}
\begin{picture}(131,114)(0,-10)
\thicklines
\path(50,35)(50,55)(30,55)
	(30,35)(50,35)
\path(100,35)(100,55)(80,55)
	(80,35)(100,35)
\path(130,35)(130,55)(110,55)
	(110,35)(130,35)
\path(10,35)(10,10)
\path(40,35)(40,10)
\dottedline{5}(10,55)(10,65)(22,71)
\dottedline{5}(40,55)(40,65)(10,80)(10,90)
\dottedline{5}(40,90)(40,80)(30,74)
\dottedline{5}(90,80)(90,55)
\dottedline{5}(120,80)(120,55)
\path(102,16)	(97.251,14.656)
	(93.855,13.302)
	(90.000,10.000)

\path(90,10)	(89.694,7.433)
	(90.000,5.000)

\path(90,5)	(90.000,3.266)
	(90.000,0.000)

\path(20,35)(20,55)(0,55)
	(0,35)(20,35)
\path(90,35)	(90.000,31.734)
	(90.000,30.000)

\path(90,30)	(89.694,27.567)
	(90.000,25.000)

\path(90,25)	(92.634,22.179)
	(96.225,20.169)
	(100.453,18.699)
	(105.000,17.500)
	(109.547,16.301)
	(113.775,14.831)
	(117.366,12.821)
	(120.000,10.000)

\path(120,10)	(120.306,7.433)
	(120.000,5.000)

\path(120,5)	(120.000,3.266)
	(120.000,0.000)

\path(120,35)	(120.000,31.734)
	(120.000,30.000)

\path(120,30)	(120.292,27.565)
	(120.000,25.000)

\path(120,25)	(116.497,21.840)
	(113.377,20.450)
	(109.000,19.000)

\put(65,43){\makebox(0,0){\Large =}}
\put(10,45){\makebox(0,0){$a$}}
\put(40,45){\makebox(0,0){$b$}}
\put(90,45){\makebox(0,0){$b$}}
\put(120,45){\makebox(0,0){$a$}}
\put(25,5){\makebox(0,0){$j$}}
\put(5,90){\makebox(0,0){$k$}}
\put(45,90){\makebox(0,0){$i$}}
\put(105,0){\makebox(0,0){$j$}}
\put(85,85){\makebox(0,0){$k$}}
\put(125,85){\makebox(0,0){$i$}}
\end{picture}
}\hspace{1in}\mbox{if $(i,j)<(k,j)$}
\]
We have drawn in solid lines $\Uq$-elements figuring in the AGS relations
associated to the $j$-th vertex. All attached to other vertices are dashed.

\item[iii)] Fields attached to the same edge form a quasi-triangular 
Hopf algebra:
\begin{enumerate}
\item $R^{-1}_{12}\stackrel{1}{U}_{(i,j)}\stackrel{2}{U}_{(i,j)}R_{12}=
                  \stackrel{2}{U}_{(i,j)}\stackrel{1}{U}_{(i,j)}$
\item $\cdot\stackrel{1}{U}_{(j,i)}\stackrel{2}{U}_{(i,j)}=1$. This property is
sometimes called the cancellation of a backtracking:

\[
\raisebox{-0.7in}{
\setlength{\unitlength}{0.0125in}
\begin{picture}(91,95)(0,-10)
\thicklines
\path(61,40)(61,20)(41,20)
	(41,40)(61,40)
\path(51,20)(51,5)
\path(95,5)(95,60)
\path(21,40)	(21.000,43.266)
	(21.000,45.000)

\path(21,45)	(20.806,47.460)
	(21.000,50.000)

\path(21,50)	(22.683,52.844)
	(25.226,55.774)
	(28.156,58.317)
	(31.000,60.000)

\path(31,60)	(33.263,60.581)
	(36.000,60.774)
	(38.737,60.581)
	(41.000,60.000)

\path(41,60)	(43.843,58.317)
	(46.774,55.774)
	(49.317,52.844)
	(51.000,50.000)

\path(51,50)	(51.194,47.460)
	(51.000,45.000)

\path(51,45)	(51.000,43.266)
	(51.000,40.000)

\path(31,40)(31,20)(11,20)
	(11,40)(31,40)
\path(21,20)	(21.000,16.734)
	(21.000,15.000)

\path(21,15)	(21.194,12.540)
	(21.000,10.000)

\path(21,10)	(18.887,7.113)
	(16.000,5.000)

\path(16,5)	(13.737,4.419)
	(11.000,4.226)
	(8.263,4.419)
	(6.000,5.000)

\path(6,5)	(3.113,7.113)
	(1.000,10.000)

\path(1,10)	(0.565,12.293)
	(0.613,15.080)
	(1.000,20.000)

\path(1,20)	(1.000,23.305)
	(1.000,27.500)
	(1.000,31.695)
	(1.000,35.000)

\path(1,35)	(1.000,38.305)
	(1.000,42.500)
	(1.000,46.695)
	(1.000,50.000)

\path(1,50)	(1.000,53.469)
	(1.000,56.235)
	(1.000,60.000)

\put(36,70){\makebox(0,0){$i$}}
\put(75,25){\makebox(0,0){\Large =}}
\put(21,30){\makebox(0,0){$a$}}
\put(51,30){\makebox(0,0){$a$}}
\put(36,0){\makebox(0,0){$j$}}
\end{picture}
}
\]
\end{enumerate}
\end{description}
\noindent
{\em Remarks:} 1)~The properties [i] and [iii] are obviously agreed with
our definition of qQCD$_2$ (see, \eg, the pictorial illustration in
Figure~1, and the discussion preceding it). The relations [ii] follow from
transformations of modules associated to vertex tangles. Of course, being
made, such a move has to be compensated somewhere by its reciprocal for a
whole construction to remain invariant.\\
2)~As a closed surface can be projected onto $R^2$ only locally, one has to
use gluing homomorphisms to assembly a Riemann surface out of flat
pieces. These homomorphisms match the gauge field algebra relations on 
different pieces and have to be added in order to complete the
construction.\\
3)~The set of the AGS relations is distinguished by an observation that they
generate a lattice Kac-Moody algebra in the sense of \rf{AFSTS}. However,
they do not constitute all possible symmetries of the qQCD$_2$ functor.

\newsection{Concluding remarks}

\begin{enumerate}

\item The first natural question to ask is whether the results of this paper
could be generalized to higher dimensions. The answer is certainly ``No''!
The reason for it is that, in dimensions bigger than 3, there is no natural
ordering of faces incident to an edge in a complex. It restricts the class
of acceptable Hopf algebras to triangular ones. Then the corresponding
construction essentially coincides with the classical Wilsonian LGT.

\item It is tempting to interpret the topological invariant considered in
this paper as some suitable generalization of \eq{fingr} and the
construction of topological qCDQ$_3$ as a generalization of $H^1(C,G)$.
Unfortunately, we are able to say nothing constructive about it. However,
in the 2-dimensional case, a notion of a quantum moduli space could
presumably be formulated \cite{FR}, which leaves some hope for the future.

\item We conjecture that qQCD$_3$ with $\Uq(su(n))$ gauge group possesses a
continuum limit equivalent to a gauge theory whose action includes both
Yang-Mills and Chern-Simons terms. One could introduce a non-zero coupling
constant in the root-of-unity case as well, which implies some deformation
of Chern-Simons theory. A meaning of this procedure is absolutely unclear
to us.

\end{enumerate}

{\Large \bf Acknowledgments}
\bigskip

I thank V.Turaev for the fruitful discussion. This work was supported by the
EEC program ``Human Capital and Mobility'' under the contract
ERBCHBICT9941621.

\end{document}